\documentclass[journal,10pt,onecolumn]{IEEEtran}
\usepackage{amssymb,amsmath,graphicx}
\usepackage{algorithmicx}
\usepackage{algpseudocode}
\usepackage{mdframed}
\usepackage{xspace}
\usepackage{graphicx}
\usepackage{graphics}
\usepackage{caption}
\usepackage{subcaption}
\usepackage{subfig}
\usepackage{multirow}
\usepackage{epstopdf}
\usepackage{array}
\usepackage{cite}
\usepackage{breqn}
\setkeys{breqn}{breakdepth={1}}

\usepackage[bordercolor=white,backgroundcolor=gray!30,linecolor=black,colorinlistoftodos]{todonotes}

\usepackage{soul}

\makeatletter
\let\old@ps@headings\ps@headings
\let\old@ps@IEEEtitlepagestyle\ps@IEEEtitlepagestyle
\def\confheader#1{%
  \def\ps@headings{%
    \old@ps@headings%
    \def\@oddhead{\strut\hfill#1\hfill\strut}%
    \def\@evenhead{\strut\hfill#1\hfill\strut}%
  }%
  \def\ps@IEEEtitlepagestyle{%
    \old@ps@IEEEtitlepagestyle%
    \def\@oddhead{\strut\hfill#1\hfill\strut}%
    \def\@evenhead{\strut\hfill#1\hfill\strut}%
  }%
  \ps@headings%
}
\makeatother

\confheader{%
  Accepted for publication in IEEE Transactions on Aerospace and Electronics Systems, June 2015
}

\newcommand{\rlsb}{RLSBE\xspace}
\newcommand{\omb}{OMBE\xspace}
\newcommand{\sfa}{SFA\xspace}
\newcommand{\fbe}{FBEA\xspace}
\newcommand{\ie}{i.e.}
\newcommand{\eg}{e.g.}
\ifCLASSINFOpdf
\else
\fi
\begin{document}
%
\title{A Practical Bias Estimation Algorithm for Multisensor--Multitarget Tracking}
%
%
%
\author{\authorblockN{Ehsan Taghavi, R. Tharmarasa and T. Kirubarajan\\}
\authorblockA{McMaster University,
Hamilton, Ontario, Canada\\
Email: $\{${taghave, tharman, kiruba}$\}$@mcmaster.ca}\\
\and
\authorblockN{Yaakov Bar-Shalom\footnote{Yaakov Bar--Shalom was supported by ARO Grant W$991$NF--$10$--$1$--$0369$.}\\}
\authorblockA{University of Connecticut,
Storrs, Connecticut, USA\\
Email: ybs@ee.uconn.edu}\\
\and \authorblockN{Mike McDonald\\} \authorblockA{Defence Research and Development Canada\\
Ottawa, Ontario, Canada\\
Email:  mike.mcdonald@drdc-rddc.gc.ca}
}
\maketitle

\begin{abstract}
Bias estimation or sensor registration is an essential step in ensuring the accuracy of global tracks in multisensor-multitarget tracking. Most previously proposed algorithms for bias estimation rely on local measurements in centralized systems or tracks in distributed systems, along with additional information like covariances, filter gains or targets of opportunity. In addition, it is generally assumed that such data are made available to the fusion center at every sampling time. In practical distributed multisensor tracking systems, where each platform sends local tracks to the fusion center, only state estimates and, perhaps, their covariances are sent to the fusion center at non-consecutive sampling instants or scans.
That is, not all the information required for exact bias estimation at the fusion center is available in practical distributed tracking systems. In this paper, a new algorithm that is capable of accurately estimating the biases even in the absence of filter gain information from local platforms is proposed for distributed tracking systems with intermittent track transmission. Through the calculation of the Posterior Cram\'er--Rao lower bound and various simulation results, it is shown that the performance of the new algorithm, which uses the tracklet idea and does not require track transmission at every sampling time or exchange of filter gains, can approach the performance of the exact bias estimation algorithm that requires local filter gains.
\end{abstract}

\begin{IEEEkeywords}
Multitarget--multisensor tracking, registration, bias estimation, tracklets, distributed fusion.
\end{IEEEkeywords}

%
\IEEEpeerreviewmaketitle

\section{Introduction}
Bias estimation and compensation are essential steps in distributed tracking systems. The objective of sensor registration is to estimate the biases in sensor  measurements, such as scaling and offset biases in range and azimuth measurements of a radar, clock bias and/or uncertainties in sensor positions \cite{bar2011tracking}.
In a distributed multisensor tracking scenario, each local tracker provides its own estimates of target states for fusion. Local filters can be, \eg, Kalman filters or Interacting Multiple Model (IMM) estimators with different motion models. These local tracks, \ie, state estimate vectors and associated covariance matrices, are sent to the fusion center for further processing. Next, the fusion center carries out track--to--track fusion. The fusion is done sequentially subsequent to the estimation of biases based on common targets that are tracked by various sensors in different locations.

Usually, bias estimation is considered as a two--sensor problem \cite{friedland1969treatment,dana1990registration} where a stacked vector is assumed with all unknown biases and states target. A drawback of this approach is the computational burden due to increased dimension of the stacked vector. In addition, most of the algorithms proposed for estimating the biases operate on the measurements directly \cite{nabaa1999solution}. That is, such methods perform filtering on the measurements received from sensors, which also include the biases. In many practical tracking systems, access to measurements before tracking at the sensor level is not always feasible. That is, sensors may provide only processed tracks to the user for further processing \cite{bar2011tracking}. Thus, methods that can simultaneously handle track--to--track fusion and bias estimation are needed.

Although there are many different methods in the literature for bias estimation and compensation, there is still a need for a method that requires only the local track estimates and associated covariance matrices for bias estimation. In \cite{okello2003maximum} and \cite{okello2004joint} a joint track-to-track bias estimation and fusion algorithm based on equivalent measurements of the local tracks was proposed. In \cite{huang2012pseudo}, another approach based on pseudo-measurement along with the Expectation-Maximization (EM) algorithm to perform joint fusion and registration was proposed. A different method that uses a multistart local search to handle the joint track-to-track association and bias estimation problem was introduced in \cite{papageorgiou2008simultaneous}. The concept of pseudo-measurement was used in \cite{lin2004exact} for \textit{exact} bias estimation with further extensions in \cite{lin2004multisensor} and \cite{lin2005multisensor}. In order to achieve exact bias estimation, the algorithms in \cite{lin2004exact,lin2004multisensor,lin2005multisensor} require the Kalman gains from local trackers,  which are not normally sent to the fusion center in practical systems \cite{drummond2002track}. Moreover, the previously mentioned algorithms assumed that the fusion center receives the local tracks from all sensors at every time step, which is not realistic in systems with bandwidth limitations \cite{chen2007track}. In addition, these methods require perfect knowledge about each local filter and its dynamic model. Also, as the number of sensors increases, the bias estimation problem suffers from the curse of dimensionality because of the commonly used stacked bias vector implementation \cite{friedland1969treatment}. Finally, as the number of sensors changes over time, the algorithms in \cite{lin2004exact,lin2004multisensor,lin2005multisensor} require appropriate pseudo--measurement to be defined for the specific number of sensors.

In this paper, these issues are addressed and a practical solution, which is mathematically sound and computationally feasible, is presented. The new approach is based on reconstructing the Kalman gains of the local trackers at the fusion center. In this approach, the tracklet method \cite{drummond2002track,drummond1997hybrid,drummond1997tracklets} along with sequential update as a fusion method is used to provide a low computational cost algorithm for bias estimation. Also, some of the constraints that were discussed above are relaxed in the proposed algorithm. The main contributions of the new algorithm are: a) reconstruction of Kalman gains at the fusion center, b) relaxing the constraints on receiving local tracks at every time step, c) correcting local tracks at the fusion center and d) providing a fused track with low computational cost.

The paper is structured as follows: The bias model and the assumptions for bias estimation are discussed in Section \ref{formulation}. In Section \ref{review}, a review of the exact bias estimation method \cite{lin2004exact,lin2004multisensor,lin2005multisensor} is given. The new approach and its mathematical developments are given in Section \ref{new}. Section \ref{pcrlb} presents the calculation of the Cram\'er--Rao Lower Bound (CRLB) for proposed algorithm. Section \ref{results} demonstrates the performance of the new algorithm for synchronous sensors and compares it with that of the method in \cite{lin2004exact} and shows comparisons with the CRLB. Conclusions are discussed in Section \ref{conclusion}.

\section{\label{formulation} Problem Formulation}
Assume that there are $M$ sensors reporting range and azimuth measurements\footnote{While this assumes $2$--D radars, the extension to $3$--D radars is straightforward.} in polar coordinates of $N$ targets in the common surveillance region. Note that $N$ is not exactly known to the algorithm and that it could be time varying. That is, bias estimation is carried out based on time--varying and possibly erroneous numbers of tracks reported by the local trackers.  The model for the measurements originating from a target with biases at time $k$ in polar coordinates (denoted by superscript $p$) for sensor $s$ is \cite{lin2004exact,lin2004multisensor,lin2005multisensor}
\begin{equation}
z_{s}^{p}(k)=\left[\begin{array}{c}
r_{s}^{p}(k)\\
\theta_{s}^{p}(k)
\end{array}\right]=\left[\begin{array}{c}
\left[1+\epsilon_{s}^{r}(k)\right]r_{s}(k)+b_{s}^{r}(k)+w_{s}^{r}(k)\\
\left[1+\epsilon_{s}^{\theta}(k)\right]\theta_{s}(k)+b_{s}^{\theta}(k)+w_{s}^{\theta}(k)
\end{array}\right] \;\;\;\;\;\; s={1,\dots,M}
\label{eq:measure}
\end{equation}
where $r_{s}(k)$ and $\theta_{s}(k)$ are the true range and azimuth, respectively, $b_{s}^{r}(k)$ and $b_{s}^{\theta}(k)$ are the offset biases in the range and azimuth, respectively, $\epsilon_{s}^{r}(k)$ and $\epsilon_{s}^{\theta}(k)$ are the scale biases in the range and azimuth, respectively. The measurement noises  $w_{s}^{r}(k)$ and $w_{s}^{\theta}(k)$ in range and bearing are zero-mean with corresponding variances $\sigma_{r}^{2}$ and $\sigma_{\theta}^{2}$, respectively, and are assumed mutually independent.

The bias vector $\beta_{s}(k)=\left[\begin{array}{cccc}
b_{s}^{r}(k) & b_{s}^{\theta}(k) & \epsilon_{s}^{r}(k) & \epsilon_{s}^{\theta}(k)\end{array}\right]^{\mathrm{T}}$ can be modeled as an unknown constant over a certain window of scans (non--random variable). Consequently, the maximum likelihood (ML) estimator \cite{balleri2007maximum} or the weighted least squares (LS) estimator \cite{asparouhov2010weighted} can be used for bias estimation. On the other hand, a Gauss-Markov random model \cite{rue2005gaussian} can also be used, in which case a Kalman filter can be adopted for bias estimation.
We model the measurement as
\begin{eqnarray}
z_{s}^{p}(k)=\left[\begin{array}{c}
r_{s}(k)\\
\theta_{s}(k)
\end{array}\right]+C_{s}(k)\beta_{s}(k)+\left[\begin{array}{c}
w_{s}^{r}(k)\\
w_{s}^{\theta}(k)
\end{array}\right]
\end{eqnarray}
where
\begin{eqnarray}
C_{s}(k)\overset{\triangle}{=}\left[\begin{array}{cccc}
1 & 0 & r_{s}(k) & 0\\
0 & 1 & 0 & \theta_{s}(k)
\end{array}\right]
\label{eq:C}
\end{eqnarray}
Here, the measured azimuth $\theta_{s}^{m}(k)$ and
range $r_{s}^{m}(k)$ can be utilized in \eqref{eq:C} without any significant
loss of performance \cite{lin2004exact,lin2004multisensor,lin2005multisensor}.

Estimating the bias vector $\beta_s(k)$ for all the sensors is the main objective of this paper. After bias estimation, all the biases can be compensated for in the state estimates at the fusion center.


Since target motion is better modeled and most trackers operate in Cartesian coordinates, the polar measurements are converted into Cartesian coordinates. It is assumed that this does not introduce biases \cite{bar2001estimation}; this is verified in the simulations. Then, sensor $s$ has the measurement equation (with the same $H_s(k)=H(k)$ for all $s$)
\begin{eqnarray}
z_{s}(k) & = & H(k)\mathbf{x}(k)+B_{s}(k)C_{s}(k)\beta_{s}(k)+w_{s}(k)\label{eq:pol2cart}
\end{eqnarray}
where the state vector $\mathbf{x}(k)=\left[\begin{array}{cccc}
x(k) & \dot{x}(k) & y(k) & \dot{y}(k)\end{array}\right]^{\mathrm{T}}$ and $H(k)$ is the measurement matrix given by
\begin{eqnarray}
H(k) & = & \left[\begin{array}{cccc}
1 & 0 & 0 & 0\\
0 & 0 & 1 & 0
\end{array}\right]\overset{\triangle}{=}H
\end{eqnarray}
Since distributed tracking systems may cover a large geographical area, the earth can no longer be assumed to be flat and coordinate transformations need to include an earth curvature model like WGS-84 \cite{li2002multi,yeddanapudi1997imm}.

The matrix $B_{s}(k)$ is a nonlinear function with the true range and azimuth as its arguments. A constant $B_{s}(k)C_{s}(k)$  also results in incomplete observability as discussed in \cite{lin2005multisensor}. Using the measured azimuth $\theta_{s}^{m}(k)$ and range $r_{s}^{m}(k)$ from sensor $s$, $B_{s}(k)$ can be written as \cite{bar2001estimation}
\begin{eqnarray}
B_{s}(k) & = & \left[\begin{array}{cc}
\cos\theta_{s}^{m}(k) & -r_{s}^{m}(k)\sin\theta_{s}^{m}(k)\\
\sin\theta_{s}^{m}(k) & r_{s}^{m}(k)\cos\theta_{s}^{m}(k)
\end{array}\right]
\end{eqnarray}
Finally, the new covariance matrix of the measurement in Cartesian coordinates (omitting index $k$ in the measurements for clarity) is given by
\begin{eqnarray}
R_{s}(k) & = & \left(\begin{array}{cc}
r_{s}^{2}\sigma_{\theta}^{2}\sin^{2}\theta_{s}+\sigma_{r}^{2}\cos^{2}\theta_{s} & \left(\sigma_{r}^{2}-r_{s}^{2}\sigma_{\theta}^{2}\right)\sin\theta_{s}\cos\theta_{s}\\
\left(\sigma_{r}^{2}-r_{s}^{2}\sigma_{\theta}^{2}\right)\sin\theta_{s}\cos\theta_{s} & r_{s}^{2}\sigma_{\theta}^{2}\cos^{2}\theta_{s}+\sigma_{r}^{2}\sin^{2}\theta_{s}
\end{array}\right)\label{eq:CartR}
\end{eqnarray}
where one can use the observed range and azimuth as well.

\section{\label{review} Review of Synchronous Sensor Registration}
In this section, the bias estimation method introduced in \cite{lin2004exact,lin2005multisensor,lin2004multisensor} for synchronous sensors with known sensor locations is reviewed. Further, the methods in our previous work \cite{taghavi2013bias} are examined in more detail and are extended in this paper with various simulations and the calculation of the lower bounds for bias estimation in multisensor--multitarget scenarios.

Consider a multisensor tracking system with the decentralized architecture \cite{bar2011tracking}. In this case, each local tracker runs its own filtering algorithm and obtains a local state estimate using only its own measurements. Then, all local trackers send their estimates to the fusion center where bias estimation is addressed. Only after bias estimation can the fusion center fuse local estimates correctly to obtain accurate global estimates.

The dynamic equation for the target state is
\begin{eqnarray}
\mathbf{x}(k+1) & = & F(k)\mathbf{x}(k)+v(k)\label{eq:dynamic}
\end{eqnarray}
where $F(k)$ is the transition matrix, and $v(k)$ is a zero-mean additive white Gaussian noise with covariance $Q(k)$.

Because the local trackers are not able to estimate the biases on their own, they yield inaccurate estimates of tracks by assuming no bias in their measurements. Hence, the state space model considered by local trackers for a specific target $t$ and sensor $s$ is
\begin{eqnarray}
\mathbf{x}^{t}(k+1) & = & F(k)\mathbf{x}^{t}(k)+v(k)\label{eq:dynamicLocal}\\
z_{s}^{t}(k) & = & H(k)\mathbf{x}^{t}(k)+w_{s}(k)\label{eq:measureLocal}
\end{eqnarray}
The difference between \eqref{eq:measure} and \eqref{eq:measureLocal} is that the latter has no bias term and, as a result, the local tracks are bias-ignorant \cite{lin2004exact,lin2005multisensor,lin2004multisensor}. Note that this mismatch should be compensated for.

\subsection{The pseudo-measurement of the bias vector}
In this subsection, a brief discussion on how to find an informative pseudo-measurement by using the local tracks for the case $M=2$ synchronized sensors is presented, based on the method given in \cite{lin2004exact,lin2005multisensor,lin2004multisensor}. As in these previous works, it is assumed that the local platforms run a Kalman filter-based tracker, although this assumption may not always be valid. However, as shown in the sequel, multiple-model based trackers can be handled within the proposed framework with some extensions.

In \cite{lin2004exact,lin2005multisensor,lin2004multisensor} it was assumed that one has access to the filter gain $W_1(k+1)$ and the residual $\nu_1(k+1)$ from the Kalman filter of local tracker $1$ \cite{welch1995introduction}. Then, one can write
\begin{eqnarray}
\hat{\mathbf{x}}_{1}(k+1\mid k+1) & = & F(k)\hat{\mathbf{x}}_{1}(k\mid k)+W_{1}(k+1)\nu_{1}(k+1)\nonumber \\
 & = & F(k)\hat{\mathbf{x}}_{1}(k\mid k)+W_{1}(k+1)\left[z_1(k+1)-\hat{z}_1^0(k+1\mid k)\right]\nonumber \\
 & = & \left[I-W_{1}(k+1)H(k+1)\right]F(k)\hat{\mathbf{x}}_{1}(k\mid k)+W_{1}(k+1)\left[H(k+1)\right.\nonumber \\
 &  & F(k)\mathbf{x}_{1}(k)+H(k+1)v(k)+B_{1}(k+1)C_{1}(k+1)\beta_{1}(k+1)\nonumber\\
 &  & \left.+w_{1}(k+1)\right]
\label{eq:pseudoProof}
\end{eqnarray}
Note that the predicted measurement $\hat{z}_1^0(k+1\mid k)$ is based on the measurement in which no bias is assumed by local tracker $1$, \ie, tracker $1$ used a bias-ignorant measurement model. Therefore, there is no term related to biases in the predicted measurement.

Hence, if the local state estimate is moved to the left--hand side of \eqref{eq:pseudoProof}, and left-multiplied by the left pseudo-inverse \cite{horn2012matrix} of the gain, one has
\begin{eqnarray}
z_{b}^{1}(k+1) & \triangleq & W_{1}^{\dagger}(k+1)\left[\hat{\mathbf{x}}_{1}(k+1\mid k+1)-\right.\left.(I-W_{1}(k+1)H(k+1))F(k)\hat{\mathbf{x}}_{1}(k\mid k)\right]\nonumber \\
 & = & H(k+1)F(k)\mathbf{x}(k)+H(k+1)v(k)+B_{1}(k+1)C_{1}(k+1)\beta_{1}(k+1) \nonumber \\
 &   & +w_{1}(k+1)
\label{eq:measure1}
\end{eqnarray}
where the pseudo-inverse of the gain is
\begin{eqnarray}
W_{s}^{\dagger} & \triangleq & \left(W_{s}^{\mathrm{T}}W_{s}\right)^{-1}W_{s}^{\mathrm{T}}\label{eq:pseudo}
\end{eqnarray}
Similarly, one can define
\begin{eqnarray}
z_{b}^{2}(k+1) & \triangleq & W_{2}^{\dagger}(k+1)\left[\hat{\mathbf{x}}_{2}(k+1\mid k+1)\right.-\left.(I-W_{2}(k+1)H(k+1))F(k)\hat{\mathbf{x}}_{2}(k\mid k)\right]\nonumber \\
 & = & H(k+1)F(k)\mathbf{x}(k)+H(k+1)v(k)+B_{2}(k+1)C_{2}(k+1)\beta_{2}(k+1) \nonumber \\
 &   & +w_{2}(k+1)
\label{eq:measure2}
\end{eqnarray}
It is worth mentioning that $\mathbf{x}(k)$ and $v(k)$ in \eqref{eq:measure1} and \eqref{eq:measure2} are the same. Thus, a pseudo-measurement of the bias vector, as in \cite{lin2004exact,lin2005multisensor,lin2004multisensor}, can be defined as follows:
\begin{eqnarray}
z_{b}(k+1) & \triangleq & z_{b}^{1}(k+1)-z_{b}^{2}(k+1)
\label{eq:pseudoMeasure}
\end{eqnarray}
for the case of using similar sensors. Then,
\begin{eqnarray}
z_{b}(k+1) & = & B_{1}(k+1)C_{1}(k+1)\beta_{1}(k+1)-B_{2}(k+1)C_{2}(k+1)\beta_{2}(k+1)\nonumber \\
 &  & +w_{1}(k+1)-w_{2}(k+1)
\label{eq:pseudoMeasuer2}
\end{eqnarray}
That is, one has the pseudo-measurement of the bias vector
\begin{eqnarray}
z_{b}(k+1) & = & \mathcal{H}(k+1)\mathbf{b}(k+1)+\tilde{w}(k+1)\label{eq:pseudoH}
\end{eqnarray}
where the pseudo-measurement matrix $\mathcal{H}$, the bias parameter vector $\mathbf{b}$ and the pseudo-measurement noise $\tilde{w}(k+1)$ are defined as
\begin{eqnarray}
\mathcal{H}(k+1) & \triangleq & [B_{1}(k+1)C_{1}(k+1),\:-B_{2}(k+1)C_{2}(k+1)]
\label{eq:H}
\end{eqnarray}
\begin{eqnarray}
\mathbf{b}(k+1) & \triangleq & \left[\begin{array}{c}
\beta_{1}(k+1)\\
\beta_{2}(k+1)
\end{array}\right]\label{eq:b}
\end{eqnarray}
and
\begin{eqnarray}
\tilde{w}(k+1) & \triangleq & w_{1}(k+1)-w_{2}(k+1)\nonumber \\
\label{eq:wtilde}
\end{eqnarray}
The bias pseudo-measurement noises $\tilde{w}$ are additive white Gaussian with zero--mean, and their covariance is
\begin{eqnarray}
\mathcal{R}(k+1) & = & R_{1}(k+1)+R_{2}(k+1)
\label{eq:R}
\end{eqnarray}
The main property of \eqref{eq:wtilde} is its whiteness, which results in a bias estimate that is exact \cite{lin2004exact,lin2005multisensor,lin2004multisensor}. In this approach, there is no approximation in deriving \eqref{eq:pseudoH}--\eqref{eq:R} unlike the methods previously proposed in \cite{kastella2000bias,shea2000precision,van1993systematic}. This was one of the main contributions of \cite{lin2004exact}.

When the measurement matrices $H_s(k)$ are the same for different local trackers, but only the second sensor has a bias, the following simplifications result:
\begin{eqnarray}
z_{b}(k+1) & = & z_{b}^{1}(k+1)-z_{b}^{2}(k+1)\label{eq:simple1}\\
\mathbf{b}(k+1) & = & \beta_{2}(k+1)\label{eq:simple2}\\
\mathcal{H}(k+1) & = & -B_{2}(k+1)C_{2}(k+1)\label{eq:simple3}\\
\tilde{w}(k+1) & = & w_{1}(k+1)-w_{2}(k+1)\label{eq:simple4}\\
\mathcal{R}(k+1) & = & R_{1}(k+1)+R_{2}(k+1).\label{eq:simple5}
\end{eqnarray}
\subsection{The recursive least square bias estimator}
If the biases are constant over a certain window of scans, one can construct a Recursive Least Square (RLS) estimator by using the pseudo-measurement equation \eqref{eq:pseudoH} \cite{lin2004exact}. The recursion of the RLS estimator has two stages: the first is to update the bias estimate recursively for different targets and the second is to update it through different time scans.

\begin{figure}[ht]
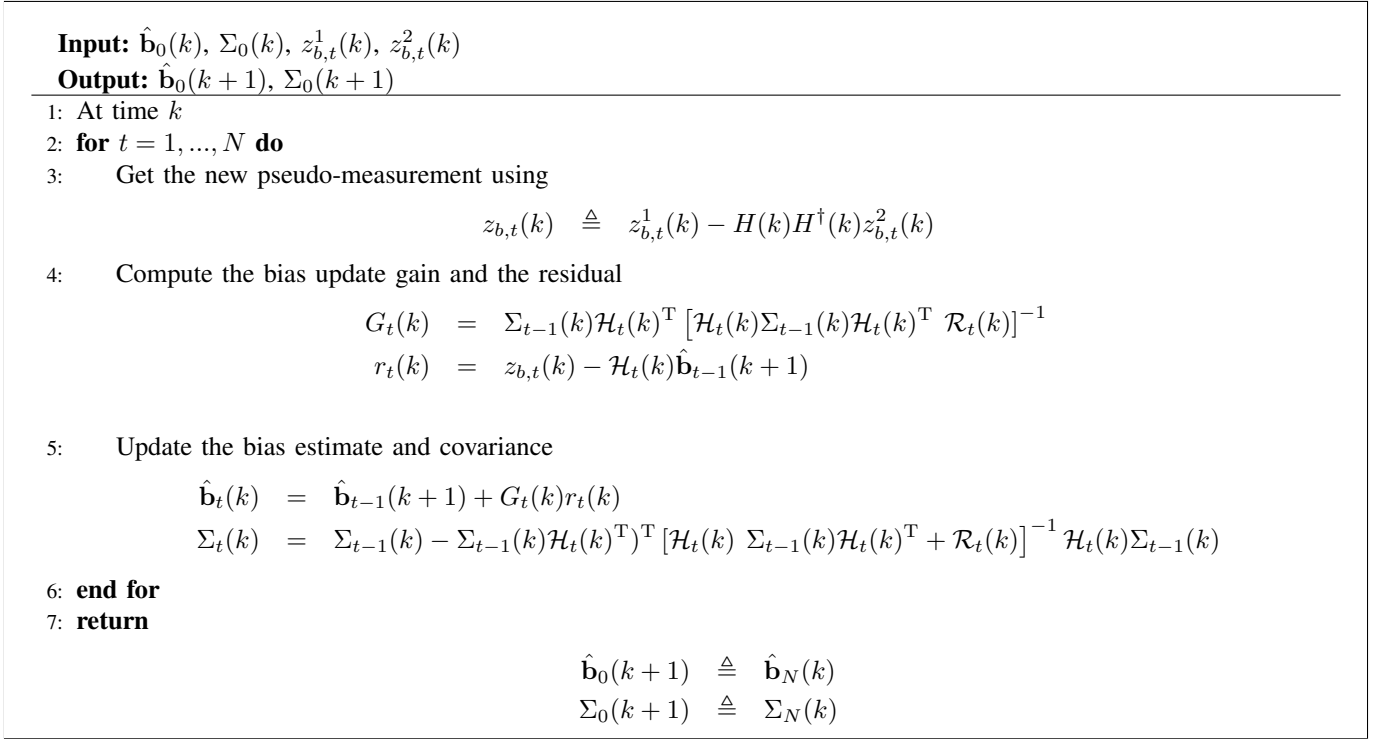

\begin{mdframed}
\begin{algorithmic}
\State {\textbf{Input:} $\hat{\mathbf{b}}_{0}(k)$, $\Sigma_{0}(k)$, $z_{b,t}^{1}(k)$, $z_{b,t}^{2}(k)$ }
\State {\textbf{Output:} $\hat{\mathbf{b}}_{0}(k+1)$, $\Sigma_{0}(k+1)$}
\hrule
\end{algorithmic}
\begin{algorithmic}[1]
\State {At time $k$}
	\For {$t=1,...,N$}
		\State {Get the new pseudo-measurement using}
		{\begin{eqnarray*}
z_{b,t}(k) & \triangleq & z_{b,t}^{1}(k)-H(k)H^{\dagger}(k)z_{b,t}^{2}(k)
\label{eq:RLSpseudoMeasure}
\end{eqnarray*}}
		\State {Compute the bias update gain and the residual}
		 {\begin{eqnarray*}
G_{t}(k) & = & \Sigma_{t-1}(k)\mathcal{H}_{t}(k)^{\mathrm{T}}\left[\mathcal{H}_{t}(k)\Sigma_{t-1}(k)\mathcal{H}_{t}(k)^{\mathrm{T}}\right.\left.\mathcal{R}_{t}(k)\right]^{-1}\nonumber \label{eq:RLSGain}\\
r_{t}(k) & = & z_{b,t}(k)-\mathcal{H}_{t}(k)\hat{\mathbf{b}}_{t-1}(k+1)\label{eq:RLSres}
\end{eqnarray*}}	
		
		\State {Update the bias estimate and covariance}
		{\begin{eqnarray*}
\hat{\mathbf{b}}_{t}(k) & = & \hat{\mathbf{b}}_{t-1}(k+1)+G_{t}(k)r_{t}(k)\label{eq:RLSb}\\
\Sigma_{t}(k) & = & \Sigma_{t-1}(k)-\Sigma_{t-1}(k)\mathcal{H}_{t}(k)^{\mathrm{T}})^{\mathrm{T}}\left[\mathcal{H}_{t}(k)\right.\left.\Sigma_{t-1}(k)\mathcal{H}_{t}(k)^{\mathrm{T}}+\mathcal{R}_{t}(k)\right]^{-1}\mathcal{H}_{t}(k)\Sigma_{t-1}(k)\nonumber \label{eq:RLSSigma}
		\end{eqnarray*}}	
	\EndFor
	\State \Return
	{\begin{eqnarray*}
\hat{\mathbf{b}}_{0}(k+1) & \triangleq & \hat{\mathbf{b}}_{N}(k)\label{eq:RLSreturn1}\\
\Sigma_{0}(k+1) & \triangleq & \Sigma_{N}(k)\label{eq:RLSreturn2}
\end{eqnarray*}}
\end{algorithmic}
\end{mdframed}
\caption{\label{fig:RLS}The Recursive Least Square Bias Estimation (\rlsb) algorithm \cite{lin2004exact}.}
\end{figure}

Assume that at time $k$, one has access to the estimate of the bias vector and its associated covariance matrix up to time $k$ as $\hat{\mathbf{b}}_{t-1}(k)$ and $\Sigma_{t-1}(k)$, form on the first $t-1$ targets and all previously updated estimates. Now, the RLS method can be carried out as in Figure \ref{fig:RLS} to update the bias estimation at time $k$ for all targets \cite{lin2004exact,lin2005multisensor,lin2004multisensor}.

Note that the covariance update equation in line $5$ of Figure \ref{fig:RLS} may cause $\Sigma_t(k)$ to lose positive definiteness due to numerical errors. To avoid this problem, the Joseph's form of the covariance update is used \cite{bar2001estimation} as
\begin{eqnarray}
\Sigma_{t}(k) & = & \left[I-G_{t}(k)\mathcal{H}_{t}(k)\right]\Sigma_{t-1}(k)\left[I-G_{t}(k)\mathcal{H}_{t}(k)\right]^{\mathrm{T}}+G_{t}(k)G_{t}(k)^{\mathrm{T}}
\label{eq:Joseph}
\end{eqnarray}
\subsection{Time-varying bias estimation: The optimal MMSE estimator}
In the case of time-varying biases with the standard linear white--Gaussian assumptions one can implement the optimal MMSE estimator based on the pseudo-measurement equation \eqref{eq:pseudoH} and the dynamic model of the bias \cite{lin2004exact}.
f
For the stacked bias vector, the dynamics model can be defines as

\begin{eqnarray}
\mathbf{b}(k+1) & = & F_{b}(k)\mathbf{b}(k)+v_{b}(k)\label{eq:dymanicBias}
\end{eqnarray}
in which $F_{b}(k)$ is the transition matrix of the stacked bias vector $\mathbf{b}$, and $v_b(k)$ is the stacked process noise of the bias vector, zero-mean white with covariance $Q_b(k)$.

Assume that at time $k$ one has access to the estimate of the bias vector and its associated covariance matrix up to time $k$ as $\hat{\mathbf{b}}_{t-1}(k\mid k)$ and $\Sigma_{t-1}(k\mid k)$, respectively. Now, a Kalman filter can be used as in Figure \ref{fig:MMSE} to update the bias estimates at time $k$ for all targets \cite{lin2004exact,lin2005multisensor,lin2004multisensor}.

\begin{figure}[htbp!]
\begin{mdframed}
\begin{algorithmic}
\State {\textbf{Input:} $\hat{\mathbf{b}}_{0}(k\mid k)$, $\Sigma_{0}(k\mid k)$, $z_{b,t}^{1}(k)$, $z_{b,t}^{2}(k)$ }
\State {\textbf{Output:} $\hat{\mathbf{b}}_{0}(k+1\mid k+1)$, $\Sigma_{0}(k+1\mid k+1)$}
\hrule
\end{algorithmic}
\begin{algorithmic}[1]
\State {At time $k$}
	\For {$t=1,...,N$}
		\State {Get the new pseudo-measurement using}
		{\begin{eqnarray*}
z_{b,t}(k) & \triangleq & z_{b,t}^{1}(k)-H(k)H^{\dagger}(k)z_{b,t}^{2}(k)
\label{eq:RLSpseudoMeasure}
\end{eqnarray*}}
		\State {Compute the bias update gain and the residual}
		 {\begin{eqnarray*}
G_{t}(k) & = & \Sigma_{t-1}(k\mid k)\mathcal{H}_{t}(k)^{\mathrm{T}}\left[\mathcal{H}_{t}(k)\Sigma_{t-1}(k\mid k)\right.\nonumber \\
 &  & \mathcal{H}_{t}(k)^{\mathrm{T}}+\left.\mathcal{R}_{t}(k)\right]^{-1}\label{eq:MMSEGain}\\
r_{t}(k) & = & z_{b,t}(k)-\mathcal{H}_{t}(k)\hat{\mathbf{b}}_{t-1}(k\mid k)\label{eq:MMSEres}
\end{eqnarray*}}	
		
		\State {Update the bias estimate and covariance}
		{\begin{eqnarray*}
\hat{\mathbf{b}}_{t}(k\mid k) & = & \hat{\mathbf{b}}_{t-1}(k\mid k)+G_{t}(k)r_{t}(k)\label{eq:MMSEb}\\
\Sigma_{t}(k\mid k) & = & \Sigma_{t-1}(k\mid k)-\Sigma_{t-1}(k\mid k)\mathcal{H}_{t}(k)^{\mathrm{T}}\nonumber \\
 &  & \left[\mathcal{H}_{t}(k)\Sigma_{t-1}(k\mid k)\mathcal{H}_{t}(k)^{\mathrm{T}}\right.\left.+\mathcal{R}_{t}(k)\right]^{-1}\mathcal{H}_{t}(k)\Sigma_{t-1}(k\mid k)\nonumber \\
\label{eq:MMSESigma}
		\end{eqnarray*}}	
	\EndFor
	\State {Update the bias estimate according to the model} {\begin{eqnarray*}
\hat{\mathbf{b}}(k+1\mid k) & \triangleq & F_{b}(k)\hat{\mathbf{b}}_{N}(k\mid k)\label{eq:MMSEreturn1}\\
\Sigma(k+1\mid k) & \triangleq & F_{b}(k)\Sigma_{N}(k\mid k)F_{b}(k)^{\mathrm{T}}+Q_{b}(k)\nonumber \\
\label{eq:MMSEreturn2}
\end{eqnarray*}}
	\State \Return
	{\begin{eqnarray*}
\hat{\mathbf{b}}_{0}(k+1\mid k+1) & = & \hat{\mathbf{b}}(k+1\mid k)\label{eq:MMSEreturn11}\\
\Sigma_{0}(k+1\mid k+1) & = & \Sigma(k+1\mid k)\label{eq:MMSEreturn22}
\end{eqnarray*}}	
\end{algorithmic}
\end{mdframed}
\caption{\label{fig:MMSE}The optimal MMSE bias estimation algorithm \cite{lin2004exact} (\omb).}
\end{figure}

\section{\label{new} The new bias estimation algorithm}
The algorithms given in Section \ref{review}, \ie, Recursive Least Square Bias Estimation (\rlsb, Figure \ref{fig:RLS}) and Optimal MMSE Bias Estimation  (\omb, Figure \ref{fig:MMSE}), are dependent on the Kalman gains provided by the local trackers, in addition to the state estimates and the associated covariance matrices at every time step. Moreover, as the number of the sensors increases, the above algorithms face an increase in computational requirements cubic in $M$. This is because a stacked vector of bias parameters is used. In addition, for $M>2$, it is challenging to extend \eqref{eq:pseudoMeasure} and \eqref{eq:pseudoH}. The extension of \eqref{eq:pseudoMeasure} and \eqref{eq:pseudoH} can be done as in \cite{bar2006multisensor} by taking $M-1$ differences. Moreover, these approaches do not address the joint fusion problem as well. With this motivation, in this section, a new approach to relax the requirement of the Kalman gain matrices availability from the local trackers is given. In addition, the new algorithm alleviates the problem of the dimensionality by taking advantage of \eqref{eq:simple1}--\eqref{eq:simple5} for a multisensor--multitarget scenario, and by solving the fusion problem as well. Finally, the algorithm is able to function properly with asynchronous local track updates.

In order to obtain the new algorithm, first,  a simple approach to calculate tracklets based on \cite{drummond2002track} is discussed. This approach makes it possible to obtain approximate equivalent measurements of the local tracks directly and efficiently without any further processing and it supports updating the bias estimates whenever a new local track is available at the fusion center. In addition, it can handle asynchronous updates from different local trackers and map them to a common time \cite{drummond2002track,drummond1997tracklets}. Then, a sequential update algorithm is proposed for the fusion step. Although it is not an optimal approach for fusing the local tracks, it is computationally cheaper than parallel update \cite{bar2011tracking}. Finally, the complete algorithm based on these two approaches with additional steps is presented.

\subsection{Equivalent measurement computation using the inverse Kalman filter method based tracklet}
The main goal in this subsection is to construct a set of approximately uncorrelated equivalent measurements (``tracklets'') from the local tracks and the associated covariance matrices for sequential update in the fusion step and also to reconstruct the local Kalman gains at the fusion center. It also relaxes the requirements of receiving the local tracks at every time step. To do so, the ``inverse Kalman filter based'' tracklet method from \cite{drummond2002track} is used (for a clear derivation and the reason for its suboptimality, see \cite[p. 577]{bar2011tracking}). Based on this method, the equations relating to the equivalent measurement vector, $\mathbf{u}_s(k, k')$, for a local track from platform $s$ at time frame $k$, given that the track data was previously sent to the global tracker for time frame $k'<k$, are as follows:
\begin{eqnarray}
\mathbf{u}_{s}(k,k') = \hat{\mathbf{x}}_{s}(k\mid k')+\mathbf{A}_{s}(k\mid k')\left[\hat{\mathbf{x}}_{s}(k\mid k)-\hat{\mathbf{x}}_{s}(k\mid k')\right]\label{eq:TrackletState}
\end{eqnarray}
where
\begin{eqnarray}
\mathbf{u}_{s}(k,k') & = & \mathbf{x}(k)+\tilde{\mathbf{u}}_{s}(k\mid k)\label{eq:Track1}\\
\mathbb{E}\left[\tilde{\mathbf{u}}_{s}(k,k')\mid\mathbf{Z}^{k'}\right] & = & 0\label{eq:Track2}\\
\mathbf{A}_{s}(k,k') & = & \mathbf{P}_{s}(k\mid k')\left[\mathbf{D}_{s}(k,k')\right]^{-1}\label{eq:Track3}\\
\mathbf{D}_{s}(k,k') & = & \mathbf{P}_{s}(k\mid k')-\mathbf{P}_{s}(k\mid k)\label{eq:Track4}
\end{eqnarray}
where $\mathbf{Z}^{k'}_{s}=\left\{\mathbf{z}_s^1,...,\mathbf{z}_s^{k'}\right\}$, and
\begin{eqnarray}
\mathbf{U}_{s}(k,k') & = & \mathbb{E}\left[\tilde{\mathbf{u}}_{s}(k,k')\left(\tilde{\mathbf{u}}_{s}(k,k)'\right)^{\mathrm{T}}\mid\mathbf{Z}^{k'}_{s}\right]\nonumber \\
 & = & \mathbf{A}_{s}(k,k')\mathbf{P}_{s}(k\mid k)\nonumber \\
 & = & \left[\mathbf{A}_{s}(k,k')-I\right]\mathbf{P}_{s}(k\mid k')\label{eq:TrackletP}
\end{eqnarray}

The information that the global tracker or the fusion center uses consists of the calculated equivalent measurement vector $\mathbf{u}_{s}(k,k')$ and its error covariance matrix $\mathbf{U}_{s}(k,k')$. Note that in order to calculate $\hat{\mathbf{x}}_{s}(k\mid k')$ and $\mathbf{P}_{s}(k\mid k')$ one needs the estimated target state $\hat{\mathbf{x}}_{s}(k'\mid k')$ and its covariance matrix $\mathbf{P}_{s}(k'\mid k')$, in addition to the dynamic models the local trackers used for filtering. Then one needs to compute $L=k-k'$ prediction steps without any new measurement data to find $\hat{\mathbf{x}}_{s}(k\mid k')$ and $\mathbf{P}_{s}(k\mid k')$. Here, it is necessary to consider the $L$-step prediction of transition and process noise covariance matrices as $F(k,k')$ and $Q(k,k')$, respectively. One can use the concept of missing observations in Kalman filter to find $F(k,k')$ and $Q(k,k')$ as in \cite[pp. 110]{durbin2012time}. It should be mentioned that all these computations require that $\mathbf{P}_{s}(k'\mid k')$, $\mathbf{P}_{s}(k\mid k')$ and $\left[\mathbf{P}_{s}(k\mid k)^{-1}-\mathbf{P}_{s}(k\mid k')^{-1}\right]$ be non-singular. This method was previously used in \cite{okello2004joint} for $k'=k-1$ (for which the non-singularity requirement does not hold in general) and with a different approach for sensor registration. For the proof of \eqref{eq:TrackletP} see Appendix \ref{appA}.

\subsection{Sequential update as fusion method}
After calculating the equivalent measurements of the state for each local track at the fusion center, they can be used as new measurements for the estimation of fused state and its covariance matrix. To do this recursively, it should be assumed that the fused state estimate and its covariance matrix at time $k'$ as $\mathbf{x}_{f}(k'\mid k')$ and $\mathbf{P}_{f}(k'\mid k')$, respectively, are already computed. For $k'=1$, the parameters $\mathbf{x}_f(k'\mid k')$ and $\mathbf{P}_f(k'\mid k')$ are initialized with $\mathbf{x}(k'\mid k')$ and $\mathbf{P}(k'\mid k')$, respectively. Then these two can be updated by following the steps in Figure \ref{fig:Sequence}. Although the sequential update is sub-optimal \cite{bar2011tracking}, it has the advantage of being computationally efficient to implement and, in addition, it is not dependent on the previous equivalent measurements at time $k'$.
\begin{figure}[htbp!]
\begin{mdframed}
\begin{algorithmic}
\State \textbf{Input:} $\mathbf{x}_{f}(k'\mid k')$ and $\mathbf{P}_{f}(k'\mid k')$, $y_{s,k}$ and $R_{s,k}$ for $s=1,...,M$
\State \textbf{Output:} $\mathbf{x}_{f}(k\mid k)$ and $\mathbf{P}_{f}(k\mid k)$
\hrule
\end{algorithmic}
\begin{algorithmic}[1]
\State \textbf{Compute}  $\mathbf{x}_{f}(k\mid k')$ and $\mathbf{P}_{f}(k\mid k')$ according to their dynamic model (prediction step)
\State \textbf{Assign:} {\begin{eqnarray*}
\mathbf{x}_{\mathrm{temp}} & = & \mathbf{x}_{f}(k\mid k')\\
\mathbf{P}_{\mathrm{temp}} & = & \mathbf{P}_{f}(k\mid k').
						\end{eqnarray*}}
	\For {$s=1,...,M$}
		\State {Update $\mathbf{x}_{\mathrm{temp}}$ and $\mathbf{P}_{\mathrm{temp}}$ with new measurement and its covariance matrix, \ie, $y_{s,k}$ and $R_{s,k}$} according to
\begin{eqnarray*}
\mathbf{x}_{\mathrm{temp}} & = & \mathbf{x}_{\mathrm{temp}}+W_{\mathrm{temp}}\tilde{\mathbf{y}}\\
W_{\mathrm{temp}} & = & \mathbf{P}_{\mathrm{temp}}H^{\mathrm{T}}\left(H\mathbf{P}_{\mathrm{temp}}H^{\mathrm{T}}+R_{s,k}\right)^{-1}\\
\tilde{\mathbf{y}} & = & y_{s,k}-H\mathbf{x}_{\mathrm{temp}}\\
\mathbf{P}_{\mathrm{temp}} & = & \left(I-W_{\mathrm{temp}}H\right)\mathbf{P}_{\mathrm{temp}}\\
H & = & \left[\begin{array}{cccc}
1 & 0 & 0 & 0\\
0 & 0 & 1 & 0
\end{array}\right]
\end{eqnarray*}
	\EndFor \\
\Return{\begin{eqnarray*}
\mathbf{x}_{f}(k\mid k) & = & \mathbf{x}_{\mathrm{temp}}\\
\mathbf{P}_{f}(k\mid k) & = & \mathbf{P}_{\mathrm{temp}}
		  \end{eqnarray*}}	 						
\end{algorithmic}
\end{mdframed}
\caption{\label{fig:Sequence}The Sequential Fusion Algorithm (\sfa) with equivalent measurements}
\end{figure}
\subsection{Multisensor fusion and track-to-track bias estimation}
The first step for implementing a general bias estimation algorithm for radar systems is to find the Kalman gains of each local track at the fusion center, by only using the state estimates and the associated covariance matrices. To do so, first, one must calculate the equivalent measurement and its covariance matrix as in \eqref{eq:TrackletState} and \eqref{eq:TrackletP} for sensor $s$, and at time frame $k$ (the target index is omitted for simplicity). Since the measurement model here is linear, one has
\begin{eqnarray}
R_{s,k} & = & H(k)\mathbf{U}_{s}(k,k')H(k)^{\mathrm{T}}\label{eq:Rstn}\\
W_{s,k} & = & \mathbf{P}_{s}(k\mid k')H(k)^{\mathrm{T}}\left[H(k)\mathbf{P}_{s}(k\mid k')H(k)^{\mathrm{T}}\right.\left.+R_{s,k}\right]^{-1}
\label{eq:Wstn}\\
y_{s,k} & = & H(k)\mathbf{u}_{s}(k,k')\label{eq:ystn}
\end{eqnarray}
Note that the reason to keep the position information only is that further in the new bias estimation algorithm we use the corrected positions as new measurements for sequential fusion. Because the equivalent measurements are used for the fused track, the Kalman gain for it (denoted by subscript $f$) at time frame $k$ can be recovered as
\begin{eqnarray}
R_{f,k} & = & H(k)\left[\sum_{i=1}^{M}\left(\mathbf{U}_{i}(k,k')\right)^{-1}\right]^{-1}H(k)^{\mathrm{T}}\label{eq:Rfn}\\
W_{f,k} & = & \mathbf{P}_{f}(k\mid k')H(k)^{\mathrm{T}}\left[H(k)\mathbf{P}_{f}(k\mid k')H(k)^{\mathrm{T}}\right.\left.+R_{f,k}\right]^{-1}
\label{eq:Wfn}
\end{eqnarray}
Note that the noises/errors in the equivalent measurements are not white so using a Kalman filter is not optimal. This amounts to the same approximation as in \cite[p. 563]{bar2011tracking}. Also in \eqref{eq:Wfn}, a common coordinate system is used for equivalent measurements. As a result, the use of a common measurement matrix $H(k)$ for all the equivalent measurements is feasible.

The sensor registration method proposed here uses the simplified formulas, \ie, \eqref{eq:simple1}--\eqref{eq:simple5}. To use these formulas, an approximately bias-compensated\footnote{Here, bias-compensated means a fused track, in which the bias-corrected equivalent measurements by using the latest estimated biases are used for fusion.} fused track needs to be found at the fusion center for the set $\left\lbrace\mathcal{S}\right\rbrace\setminus s$, where $\mathcal{S}={1,2,...,M}$. The notation $\left\lbrace\mathcal{S}\right\rbrace\setminus s$ stands for the set that contains all those elements of $\mathcal{S}$ excluding element $s$. Then, the bias estimation problem can be reduced to the case of two sensors. The first one is an equivalent sensor with fusion of bias-corrected measurements of the set $\left\lbrace\mathcal{S}\right\rbrace\setminus s$, and the second is sensor $s$ which has bias. Then the bias in sensor $s$ can be found with either the \rlsb (see Figure \ref{fig:RLS}) or \omb (see Figure \ref{fig:MMSE}) algorithm.

The next step in this approach is to correct the biases in the measurement domain of the sensors (except for sensor $s$) and then fuse them together. This can be done by going from the Cartesian coordinate of equivalent measurements to the polar coordinate of the radar and correct with the previously estimated biases. Then by correcting the covariance matrix of the new bias compensated measurements, they can be fused by sequentially updating the fused track (excluding the track from sensor $s$) by using the Sequential Fusion Algorithm (\sfa). Then, the Kalman gain for the fused and the now-corrected track can be calculated using \eqref{eq:Wfn}.

For the equivalent measurement, define
\begin{eqnarray}
\mathbf{u}_{s} & \triangleq & \left[\begin{array}{cccc}
u_{x} & \dot{u}_{x} & u_{y} & \dot{u}_{y}\end{array}\right]^{\mathrm{T}}\label{eq:udefine}
\end{eqnarray}
where time and target indexes are omitted for simplicity.
Then, to correct the biases in the measurement domain, assuming that the latest--estimated biases are $\hat{\epsilon}_{\theta}$, $\hat{\epsilon}_{r}$,  $\hat{b}_{\theta}$ and $\hat{b}_{r}$, one has\footnote{Here, the superscript ``$\operatorname{b-c}$" is used to denote the bias-corrected bearing and range.}
\begin{eqnarray}
\theta_{s}^{\operatorname{b-c}} & = & \frac{\arctan\left(\frac{u_{y}}{u_{x}}\right)-\hat{b}_{\theta}}{\left(1+\hat{\epsilon}_{\theta}\right)}\label{eq:thetafree}\\
r_{s}^{\operatorname{b-c}} & = & \frac{\sqrt{\left(u_{x}\right)^{2}+\left(u_{y}\right)^{2}}-\hat{b}_{r}}{\left(1+\hat{\epsilon}_{r}\right)}\label{eq:rfree}
\end{eqnarray}
Now that the scale and offset biases are compensated for, one can go back to Cartesian coordinates as follows:
\begin{eqnarray}
u_{x}^{\operatorname{b-c}} & = & \lambda_{\theta}r_{s}^{\operatorname{b-c}}\cos\left(\theta_{s}^{\operatorname{b-c}}\right)\label{eq:uxfree}\\
u_{y}^{\operatorname{b-c}} & = & \lambda_{\theta}r_{s}^{\operatorname{b-c}}\sin\left(\theta_{s}^{\operatorname{b-c}}\right)\label{eq:uyfree}\\
\mathcal{Y}^{\operatorname{b-c}} & = & \left[\begin{array}{cc}
u_{x}^{\operatorname{b-c}} & u_{y}^{\operatorname{b-c}}\end{array}\right]^{\mathrm{T}}\label{eq:vectorfree}
\end{eqnarray}
where
\begin{eqnarray}
\lambda_{\theta} & = & \exp\left(-\frac{\sigma_{\theta}^{2}}{2}\right)
\end{eqnarray}
is the compensation factor for the bias in coordinate conversion from polar to Cartesian \cite{longbin1998unbiased}. The method from \cite{longbin1998unbiased} is used here because of the fact that bias correction and compensation along with the changes in the covariance matrices and uncertainties may violate the assumptions made for the debaised conversion in Section \ref{formulation}. The next step is to update the covariance matrix of the corrected equivalent measurements. In addition to the term \eqref{eq:Rstn}, the additional uncertainty in the bias estimates, \ie, their associated covariance matrix and the uncertainty in the model of the radar, both in Cartesian coordinates, must be accounted for. The final formula for the covariance matrix with proper conversion from polar to Cartesian coordinate is
\begin{eqnarray}
R_{s}^{\operatorname{b-c}} & = & H(k)\mathbf{U}_{s}(k\mid k)H(k)^{\mathrm{T}}+B_{s}^{\operatorname{b-c}}(k)\left[\begin{array}{cc}
\sigma_{r}^{2} & 0\\
0 & \sigma_{\theta}^{2}
\end{array}\right]B_{s}^{\operatorname{b-c}}(k)^{\mathrm{T}}\nonumber \\
& &+K_{s}^{\operatorname{b-c}}(k)\mathbf{P}_{b,s}(k\mid k)K_{s}^{\operatorname{b-c}}(k)^{\mathrm{T}}
\label{eq:Rsfree}
\end{eqnarray}
where
\begin{eqnarray}
K_{s}^{\operatorname{b-c}}(k)&=&B_{s}^{\operatorname{b-c}}(k)C_{s}^{\operatorname{b-c}}(k)
\end{eqnarray}
and $\mathbf{P}_{b,s}(k\mid k)$ is the latest-updated bias covariance matrix at time $k$ and for sensor $s$.
Now that all the required formulations and variables are available, the new algorithm to find the bias estimates for all the sensors is given in Figure \ref{fig:NewAlg}.
\begin{figure}[htbp!]
\begin{mdframed}
\begin{algorithmic}
\State \textbf{Input:} inputs of \sfa and \rlsb (defined within the corresponding algorithms).
\State \textbf{Output:} $\mathbf{b}_{s}(k)$ and $\Sigma_{s}(k)$ for $s=1,...,M$.
\hrule
\end{algorithmic}
\begin{algorithmic}[1]
\State At time $k$
	\For {$s=1,...,M$}
		\State \textbf{Compute} $W_{s,k}$ as in \eqref{eq:Wstn} and\\
		\begin{eqnarray*}
z_{b,t}^{s}(k) & \triangleq & W_{s,k}^{\dagger}\left[\hat{\mathbf{x}}(k\mid k)-(I-W_{s,k}H_{s}(k))F(k,k-L)\hat{\mathbf{x}}(k-L\mid k-L)\right]
		\end{eqnarray*}
		\State {$\bar{s}\in \lbrace 1,...,M \rbrace \backslash s$}
			\State \textbf{Call} \sfa with inputs $\mathbf{x}_{f}^{s}(k'\mid k')$ and $\mathbf{P}_{f}^{s}(k'\mid k')$, $\mathcal{Y}_{\bar{s}}^{\operatorname{b-c}}(k\mid k)$ and ${R}_{\bar{s}}^{\operatorname{b-c}}(k\mid k)$.
			\State \textbf{Compute} $W_{f,k}^{s}$ as in \eqref{eq:Wfn} and
			\begin{eqnarray*}
z_{b,f}^{s}(k) & \triangleq & \left(W_{f,k}^{s}\right)^{\dagger}\left[\hat{\mathbf{x}}_{f}^{s}(k\mid k)-(I-W_{f,k}^{s}H_{f}(k))F(k,k-L)\hat{\mathbf{x}}_{f}^{s}(k-L\mid k-L)\right]
			\end{eqnarray*}
			\State \textbf{Call} \rlsb with inputs $z_{b,t}^{s}(k)$, $z_{b,f}^{s}(k)$ and the last update of the estimated biases and their associated covariance matrix.
		\State \Return {$\mathbf{b}_{s}(k)$ and $\Sigma_{s}(k)$}
	\EndFor	 						
\end{algorithmic}
\end{mdframed}
\caption{\label{fig:NewAlg}The Fused Bias Estimation algorithm (\fbe)}
\end{figure}

As shown in Figure \ref{fig:NewAlg}, one only needs to call \sfa and \rlsb with new input parameters. One of the advantages of this approach is that in each ``\textbf{for loop}" only a low dimensional Kalman filter that is \textit{independent} of the size of the stacked bias vector and number of the sensors is needed. In addition, the fusion of local tracks can be done by only adding one sequential update for the latest corrected measurement of sensor  $s$ to the previously fused track. It is also important to note that the there is no constraint on the rate of receiving local tracks from the individual sensors. To show how well this new algorithm performs, in the next section, the simulation results on two different scenarios are used to compare its performance with those of the previously proposed algorithm in \cite{lin2004exact,lin2004multisensor,lin2005multisensor} for synchronous sensors.

\begin{figure}[htbp!]
\centering
\includegraphics[width=3.5in]{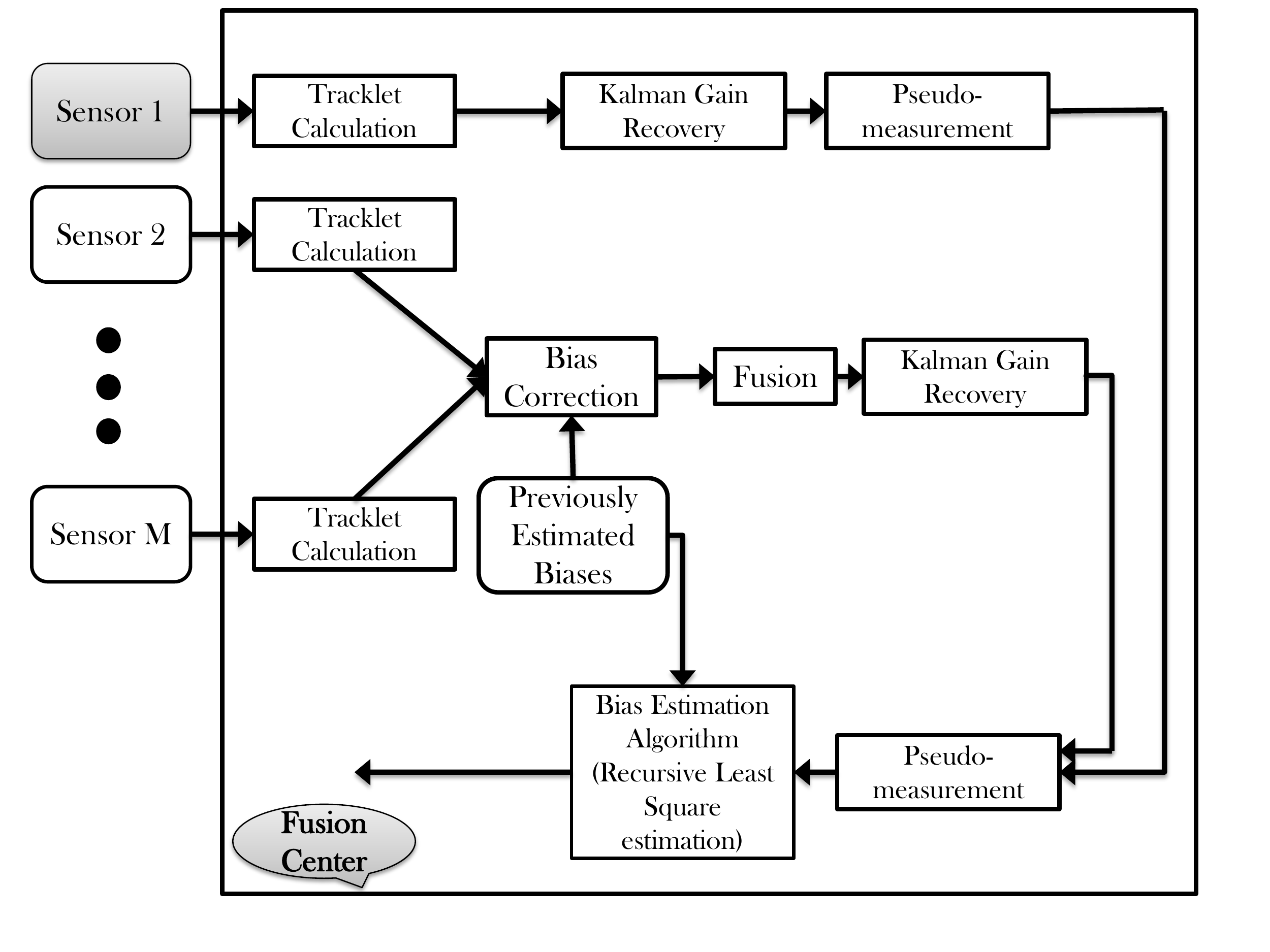}
\caption{Block diagram of the new offset and scaling bias estimation algorithm.}
\label{fig:Blkdiag}
\end{figure}

To better illustrate how the new bias estimation algorithm (\fbe) works, a block--diagram representation of the method is shown in Figure \ref{fig:Blkdiag} for a single time step estimation of the biases for the first sensor. In Figure \ref{fig:Blkdiag}, by receiving the local track estimates from all available sensors at time step $k$, the first step is to calculate the tracklet for all of them using \eqref{eq:TrackletState}--\eqref{eq:TrackletP}. Then, the equivalent measurement of the first sensor is sent for Kalman gain recovery using \eqref{eq:Rstn} and \eqref{eq:Wstn}. At the same time, the equivalent measurements of all other sensors are sent to bias correction to first remove the bias from the equivalent measurements by using the previously estimated biases at time step $k-1$ using \eqref{eq:thetafree}--\eqref{eq:vectorfree} and the Kalman gain recovery in \eqref{eq:Rfn} and \eqref{eq:Wfn}. The next step is to fuse the tracks by using \sfa algorithm. Then, the fused and corrected estimate is sent to the pseudo--measurement calculation block for each individual sensor. At this point one has a two--sensor problem with only one sensor having biases in the measurement. The output is now sent to the \rlsb algorithm along with the previously estimated biases for the first sensor so that the bias estimates can be updated at time step $k$ before proceeding to the next time step. 

\section{\label{pcrlb} Cram\'er-Rao Lower Bound for Sensor Registration}
In this section, a step-by-step procedure is given for calculating the CRLB for sensor registration algorithms as the benchmark. Rewriting \eqref{eq:pol2cart} for the case of two similar sensors, one has
\begin{equation}
z_{1}(k)-H(k)\mathbf{x}(k)=B_{1}(k)\beta_{1}(k)+w_{1}(k)
\label{eq:pcrlb_measCons11}
\end{equation}
and
\begin{equation}
z_{2}(k)-H(k)\mathbf{x}(k)=B_{2}(k)\beta_{2}(k)+w_{2}(k)
\label{eq:pcrlb_measCons12}
\end{equation}
If no biases exist, the sensors must point to the same position of the observed target. Consequently, one has
\begin{equation}
z_{1}(k)-B_{1}(k)\beta_{1}(k)-w_{1}(k)=z_{2}(k)-B_{2}(k)\beta_{2}(k)-w_{2}(k)
\label{eq:pcrlb_measCons2}
\end{equation}
and by reordering the measurement terms and using the matrix form, one can rewrite it as 
\begin{equation}
\left[z_{1}(k)-z_{2}(k)\right]=\left[\begin{array}{cc}
B_{1}(k) & -B_{2}(k)\end{array}\right]\left[\begin{array}{c}
\beta_{1}(k)\\
\beta_{2}(k)
\end{array}\right]+w_{1,2}(k)
\label{eq:pcrlb_measCons3}
\end{equation}
Further, for future use, one can denote the terms in \eqref{eq:pcrlb_measCons3} as
\begin{equation}
Y(k)=B(k)\mathbf{b}(k)+w_{1,2}(k)
\label{eq:measCons4}
\end{equation}
where
\begin{eqnarray} 
Y(k)&=&\left[z_{1}(k)-z_{2}(k)\right]\\
B(k)&=&\left[\begin{array}{cc}B_{1}(k) & -B_{2}(k)\end{array}\right]\\
\mathbf{b}(k)&=&\left[\begin{array}{c}\beta_{1}(k)\\\beta_{2}(k)\end{array}\right]
\end{eqnarray}
and $w_{1,2}(k)$ is additive white Gaussian noise with covariance matrix equal to $R_1(k)+R_2(k)$

\subsection{Calculation of the CRLB}
In the case of having two sensors and multiple targets, the CRLB can be calculated as a batch process. Taking all the (linearly independent) $K$ pairs of measurements for $N$ targets in the surveillance region, one can write the measurement equation as
\begin{equation}
\mathbf{Y}=\mathbf{g}\mathbf{b}+\mathbf{u}
\label{eq:stackedMeas}
\end{equation}
where $\mathbf{Y}$, $\mathbf{g}$ and $\mathbf{u}$ are stacked vectors given by 
\begin{equation}
\mathbf{Y}=\left[\begin{array}{ccccccc}
\left(Y^{1}(1)\right)^{\mathrm{T}} & \cdots & \left(Y^{N}(1)\right)^{\mathrm{T}} & \cdots & \left(Y^{1}(K)\right)^{\mathrm{T}} & \cdots & \left(Y^{N}(K)\right)^{\mathrm{T}}\end{array}\right]^{\mathrm{T}}
\label{eq:stackedY}
\end{equation}

\begin{equation}
\mathbf{g}=\left[\begin{array}{ccccccc}
\left(B^{1}(1)\right)^{\mathrm{T}} & \cdots & \left(B^{N}(1)\right)^{\mathrm{T}} & \cdots & \left(B^{1}(K)\right)^{\mathrm{T}} & \cdots & \left(B^{N}(K)\right)^{\mathrm{T}}\end{array}\right]^{\mathrm{T}}
\label{eq:stackedB}
\end{equation}
and
\begin{equation}
\mathbf{u}=\left[\begin{array}{ccccccc}
\left(w_{1,2}^{1}(1)\right)^{\mathrm{T}} & \cdots & \left(w_{1,2}^{N}(1)\right)^{\mathrm{T}} & \cdots & \left(w_{1,2}^{1}(K)\right)^{\mathrm{T}} & \cdots & \left(w_{1,2}^{N}(K)\right)^{\mathrm{T}}\end{array}\right]^{\mathrm{T}}
\label{eq:stackedw}
\end{equation}
Further, the covariance matrix of the noise vector $\mathbf{u}$ is
\begin{equation}
\mathcal{R}=\mathrm{diag}\left(\left[\begin{array}{ccccccc}
\mathcal{R}^{1}(1) & \cdots & \mathcal{R}^{N}(1) & \cdots & \mathcal{R}^{1}(K) & \cdots & \mathcal{R}^{N}(K)\end{array}\right]\right)
\label{eq:stackedR}
\end{equation}
where $\mathcal{R}^{i}(k)={R}_1^{i}(k)+{R}_2^{i}(k)$ and the lower index indicates a specific sensor.
As stated in \cite{bar2001estimation}, the covariance matrix of an unbiased estimator $\hat{\mathbf{b}}$ is bounded from below as
\begin{equation}
\mathbb{E}\left\{ \left(\hat{\mathbf{b}}-\mathbf{b}\right)\left(\hat{\mathbf{b}}-\mathbf{b}\right)^{\mathrm{T}}\right\} \geq\mathbf{J}^{-1}
\label{eq:lowerBound}
\end{equation}
In the above, $\mathbf{J}$ is the Fisher Information Matrix (FIM) given by
\begin{eqnarray}
\mathbf{J} & = & \mathbb{E}\left\{ \left[\nabla_{\mathbf{b}}\ln p(\mathbf{Y}\mid\mathbf{b})\right]\left[\nabla_{\mathbf{b}}\ln p(\mathbf{Y}\mid\mathbf{b})\right]^{\mathrm{T}}\right\} \mid_{\mathbf{b}=\mathbf{b}_{\mathrm{true}}}\nonumber \\
 & = & \mathbb{E}\left\{ \left[\nabla_{\mathbf{b}}\lambda\right]\left[\nabla_{\mathbf{b}}\lambda\right]^{\mathrm{T}}\right\} \mid_{\mathbf{b}=\mathbf{b}_{\mathrm{true}}}
\label{eq:J}
\end{eqnarray}
where $\mathbf{b}_{\mathrm{true}}$ is the true value of the bias vector $\mathbf{b}$, $p(\mathbf{Y}\mid\mathbf{b})$ is the likelihood function of $\mathbf{b}$, $\lambda=-\ln p(\mathbf{Y}\mid\mathbf{b})$ and $\nabla$ is gradient operator.
From \eqref{eq:stackedMeas}, one has
\begin{equation}
 p(Y\mid\mathbf{b})=\frac{1}{\left(2\pi\right)^{K}\sqrt{\left|\mathcal{R}\right|}}\exp\left\{ -\frac{1}{2}\left[\mathbf{Y}-\mathbf{g}\mathbf{b}\right]^{\mathrm{T}}\mathcal{R}^{-1}\left[\mathbf{Y}-\mathbf{g}\mathbf{b}\right]\right\}
\label{eq:logEXP}
\end{equation}
and therefore
\begin{eqnarray}
\lambda&=&\mathrm{Const.}+\frac{1}{2}\left[\mathbf{Y}-\mathbf{g}\mathbf{b}\right]^{\mathrm{T}}\mathcal{R}^{-1}\left[\mathbf{Y}-\mathbf{g}\mathbf{b}\right]
\end{eqnarray}
Using the results from \cite{graybill1976theory} to simplify the differentiations, $\nabla_{\mathbf{b}}\lambda$ can be written as
\begin{equation}
\nabla_{\mathbf{b}}\lambda=\mathbf{g}^{\mathrm{T}}\mathcal{R}^{-1}\left(\mathbf{Y}-\mathbf{g}\mathbf{b}\right)
\label{eq:derivativeLambda}
\end{equation}
which yields
\begin{equation}
\mathbf{J}=\mathbf{g}^{\mathrm{T}}\mathcal{R}^{-1}\mathbf{g}
\label{eq:JFinal}
\end{equation}
Finally, when calculating the FIM, CRLB of desired elements (here the diagonal) will be
\begin{equation}
\mathrm{CRLB}\left\{ \left[\mathbf{b}\right]_{i}\right\} =\left[\mathbf{J}^{-1}\right]_{ii}
\label{eq:stackedCRLB}
\end{equation}
for $i=1,2,3,4$.

\subsection{Multitarget--multisensor CRLB}
When number of the targets is greater than two, the same procedure as the proposed bias estimation algorithm to calculate the CRLB can be used. In this case, one should fuse all the measurements, except for sensor ``$i$'' to create a single pseudo-measurement for this ``combined'' sensor and then treat the problem as a two--sensor problem. Starting with the calculation of the combined measurement and covariance matrix of the set ${S}\i$ as in \cite{bar2011tracking} yields
\begin{eqnarray}
Z_{\mathrm{comb}} & = & \mathcal{R}_{\mathrm{comb}}\left(\sum_{j\in\{S\}\backslash i}\left(\mathcal{R}_{j}\right)^{-1}z_{j}\right)
\label{eq:Zcoupled}
\end{eqnarray}
and
\begin{eqnarray}
\mathcal{R}_{\mathrm{comb}} & = & \left(\sum_{j\in\{S\}\backslash i}\mathcal{R}_{j}^{-1}\right)^{-1}
\label{eq:Rcoupled}
\end{eqnarray}
As in the bias estimation algorithm, it is assumed that only sensor ``$i$'' has bias. This means that in the calculation of CRLB one should have access to the measurements both with and without bias. The next is to calculate $B_{\mathrm{comb}}(k)$. Similar to the bias estimation algorithm, one should define $H(k)=-B_{\mathrm{comb}}(k)$ and use the combined covariance and measurement matrices as a bias free measurement to find the CRLB of the bias estimation for sensor ``$i$''. Finally, $\mathcal{R}(k)$ can be calculated as
\begin{eqnarray}
\mathcal{R}(k) & = & \mathcal{R}_{\mathrm{comb}}(k)+\mathcal{R}_{i}(k)
\label{eq:RfinalCRLB}
\end{eqnarray}
By using \eqref{eq:Zcoupled}, \eqref{eq:Rcoupled} and \eqref{eq:RfinalCRLB}, the formulas for two sensors and multiple targets can be modified to calculate the CRLB for the case of multisensor--multitarget scenario.

\section{\label{results} Simulation results}

\subsection{\label{parameters} Motion dynamics and measurement parameters}
\begin{figure}[htbp!]
\centering
\includegraphics[width=3.5in]{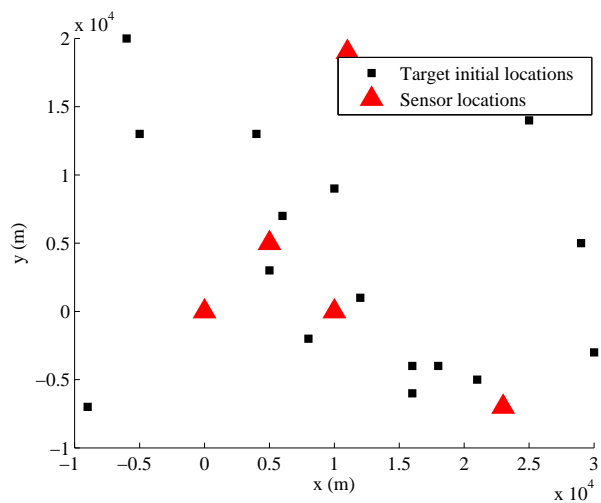}
\caption{Initial locations of the targets and sensors}
\label{fig:Scenario}
\end{figure}

Here, a distributed tracking scenario with five sensors and sixteen targets is considered as shown in Figure \ref{fig:Scenario}.
It is assumed that all sensors are synchronized. Without loss of generality and to easily compare the results of bias estimation for different sensors, all the biases are assumed as
\begin{equation}
\beta_{s}=[\begin{array}{cc}
20\mathrm{m} & 1\mathrm{mrad}\end{array}]^{\mathrm{T}}
\end{equation}
The standard deviations of measurement noises are $\sigma_{r}=10\;\mathrm{m}$ and $\sigma_{\theta}=1\;\mathrm{mrad}$ for target range and azimuth measurements, respectively. In this problem $r\simeq20000$. According to \cite{bar2001estimation}, $\frac{r\sigma_{\theta}^{2}}{\sigma_{r}}=\frac{2\times10^{4}\times10^{-6}}{10}=2\times10^{-3}\ll0.4$ which is the threshold for polar to Cartesian conversion to be unbiased. Note that this condition holds for $r\simeq200\mathrm{km}$ as well. 

Scaling biases can be handled by the proposed method as well but ignored for simplicity. The true dynamics of the targets are modeled using the Discretized Continuous White Noise Acceleration (DCWNA) or nearly constant velocity (NCV) model with $q_{x}=q_{y}=0.1\frac{\mathrm{m}^{2}}{\mathrm{s}^{3}}$ and constant turn rate model with rate $\omega=0.1\frac{\mathrm{deg}}{\mathrm{s}}$. As for the filtering in local trackers, DCWNA and Continuous Wiener Process Acceleration (CWPA) are used with various settings to be able to create different scenarios for the simulation (for detail see \cite[p. 268 and p. 467]{bar2001estimation}).





\subsection{\label{twoSensorCase} A two-sensor problem}
To compare the results of the new approach with those of the previously proposed algorithm \cite{lin2004exact,lin2004multisensor,lin2005multisensor}, the case of two sensors (placed at $(0\mathrm{m},0\mathrm{m})$ and $(5000\mathrm{m},0\mathrm{m})$ in Cartesian coordinates) is considered with the same kinematics and filter settings as in \cite{lin2004exact}. The new algorithm that uses the reconstructed Kalman gains detailed in \eqref{eq:Rstn} and \eqref{eq:Wstn} is denoted as EXL while the previous algorithm in \cite{lin2004exact,lin2004multisensor,lin2005multisensor} is denoted as EX. Note that there is no fusion step in this case and the two methods only differ in the Kalman gain calculation, which in the EXL \footnote{The EXL algorithm has the luxury of operating without the local Kalman gains.} case is reconstructed approximately.

The settings of the variables are as follows. To handle the $L=1$ and non--singularity problem, the ``tracklet with decorrelated state estimate'' \footnote{This is equivalent to the information matrix fusion method, which, for $L=1$, is algebraically equivalent to the Kalman filter (see \cite[eq. (8.4.1-14)]{bar2011tracking}).} that can be used with only one measurement in the tracklet interval \cite{drummond2002track} must be selected instead of ``tracklet computed using inverse Kalman filter'' . The sampling intervals are $T=1\mathrm{s}$. The lag between each update at the fusion center is $L=k-k'=1$. The initial state estimates are the converted measurements from polar coordinate to Cartesian coordinate with zero velocity and covariance matrix \cite{lin2004exact,lin2004multisensor,lin2005multisensor}
\begin{eqnarray}
P_{s}(0\mid0)=\mathrm{diag}\left[\begin{array}{cccc}
\left(200\mathrm{m}\right)^{2} & \left(20\mathrm{m}/\mathrm{s}^{2}\right)^{2} & \left(200\mathrm{m}\right)^{2} & \left(20\mathrm{m}/\mathrm{s}^{2}\right)^{2}\end{array}\right]
\end{eqnarray}
Finally, the initial bias parameter estimate of all the sensors are zero with initial bias covariance 
\begin{eqnarray}
\Sigma_{s}(0\mid0)=\mathrm{diag}\left[\begin{array}{cc}
\left(20\mathrm{m}\right)^{2} & \left(1\mathrm{mrad}\right)^{2}\end{array}\right]\hspace{0.5in}s=1,2
\end{eqnarray} 

\begin{figure}[htpb!]
\centering
\includegraphics[width=3.5in]{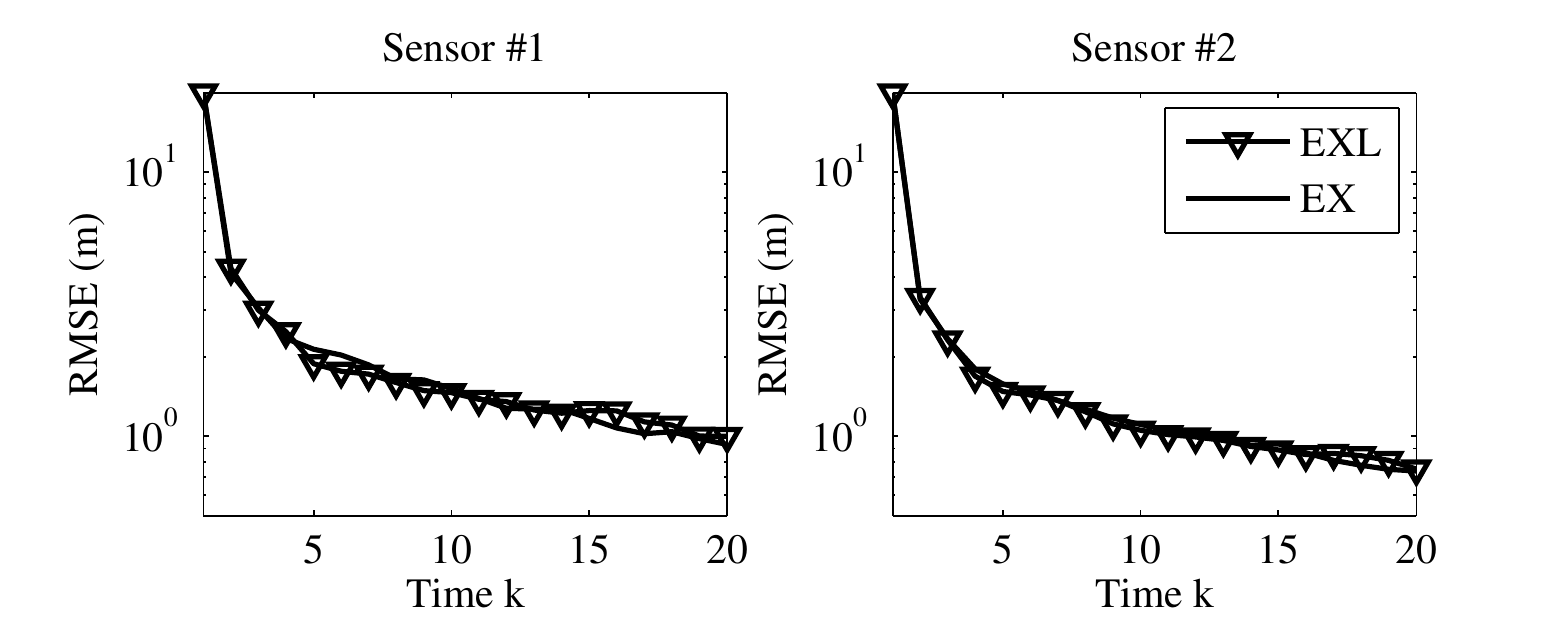}
\caption{RMSE of the bias parameter $b_{r}$ for sensors $1$ and $2$ in logarithmic scale (comparison between the previous (EX) and the proposed (EXL) algorithm)}
\label{fig:biasRMSE1}
\end{figure}

\begin{figure}[htpb!]
\centering
\includegraphics[width=3.5in]{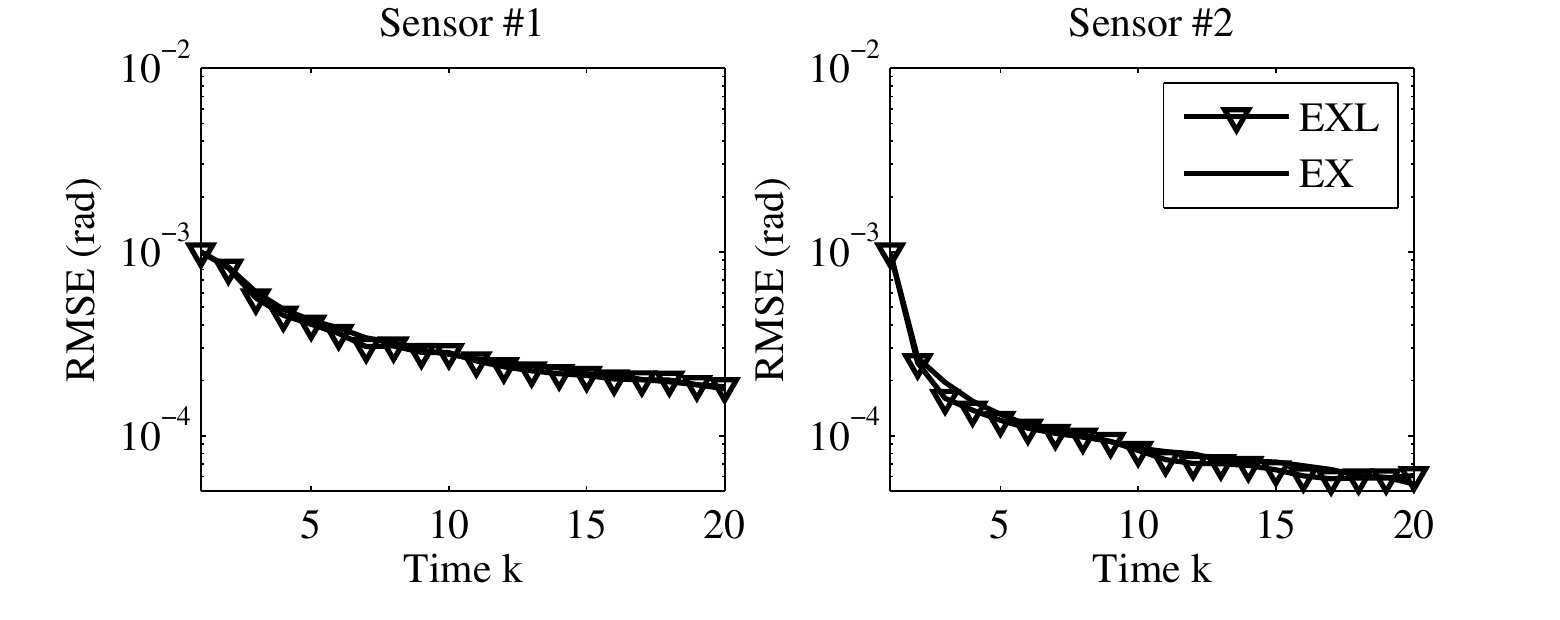}
\caption{RMSE of the bias parameter $b_{\theta}$ for sensors 1 and 2 in logarithmic scale (comparison between the previous (EX) and the proposed (EXL) algorithm))}
\label{fig:biasRMSE2}
\end{figure}

In the simulations, $100$ Monte Carlo runs are used over $20$ frames. The results of the Root Mean Squared Error (RMSE) in logarithmic scale for offset bias estimates are shown in Figures \ref{fig:biasRMSE1} and \ref{fig:biasRMSE2}. From Figures \ref{fig:biasRMSE1} and \ref{fig:biasRMSE2}, it can be seen that the performance of the EXL method, which recovers the Kalman gain at the fusion center by taking advantage of tracklet calculation, is very close to the accuracy of the EX method. The small variations in the results are due to the fact that the logarithmic scale is used to show the convergence rates clearly.

The CPU time for one iteration of bias estimation for a single target is $4.84\times 10^{-4}$ s for the EX method and $6.02\times 10^{-4}$ s for the EXL method, which represents a $20\%$ increase in CPU time for the Kalman gain reconstruction. The CPU time over all iterations and targets for bias estimation with the above methods are $0.1526$ and $0.1877$ s, respectively. All simulations are done on a computer with Intel\textsuperscript{\textregistered} Core\textsuperscript{\texttrademark} i7-3720Qm 2.60GHz processor and 8GB RAM.

\subsection{\label{fiveSensorKalman} A five--sensor problem with nearly constant velocity (NCV) Kalman filter as local tracker}
\begin{figure}[t!]
\centering
\includegraphics[width=3.5in]{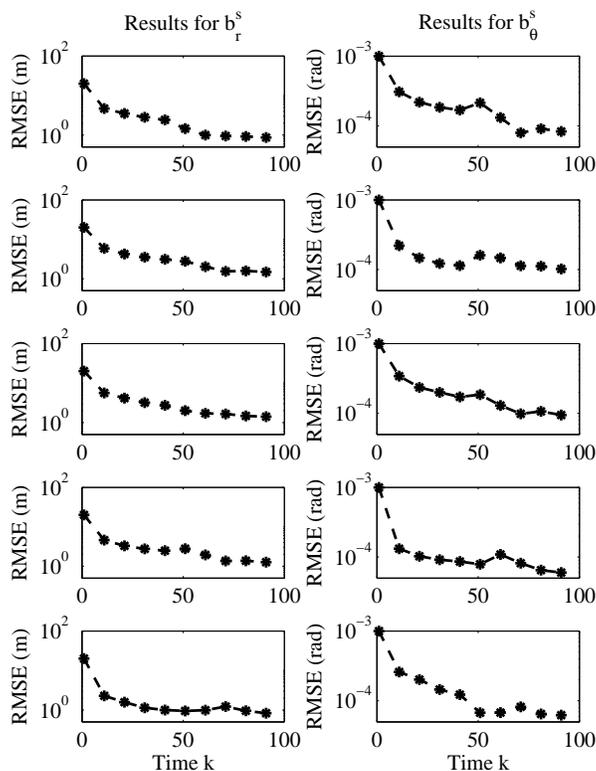}
\caption{RMSE of the bias parameter $b_{r}$ (left column) and $b_{\theta}$ (right column) for all $5$ sensors from sensor $1$ (top) to sensor $5$ (bottom) in logarithmic scale. Note that residual bias RMSE is an order of magnitude below the measurement noise standard deviations, \ie, it becomes negligible.}
\label{fig:biasRMSE}
\end{figure}

In  the second case, a scenario with five sensors with the same sixteen targets,  $L=k-k'=10$ and $k=1,...,100$ is simulated. The motivation for using $100$ time steps is to have enough update steps to demonstrate the convergence results in terms of RMSE. Here the Fused Bias Estimation Algorithm (\fbe, Figure \ref{fig:NewAlg}) is used to estimate the biases at the fusion center. The results of this simulation are shown in Figure \ref{fig:biasRMSE}.

From Figure \ref{fig:biasRMSE}, it can be seen that the  proposed algorithm works well in a scenario with five sensors and with tracklet update at every $L=10$ steps. Note that the \fbe is using only a two-dimensional state space for the bias estimation step for each sensor. For the same scenario, the previous EX algorithm would required a ten dimensional state space model. At its core, \fbe is a recursive least square (RLS) estimator. The computational complexity of RLS is of the order of $O(n^2)$, where $n$ is number of parameters to be estimated. With $n=10$ in the simulation, the computational complexity of the EX algorithm will be substantially higher.

\begin{figure}[htbp!]
\centering
\includegraphics[width=3.5in]{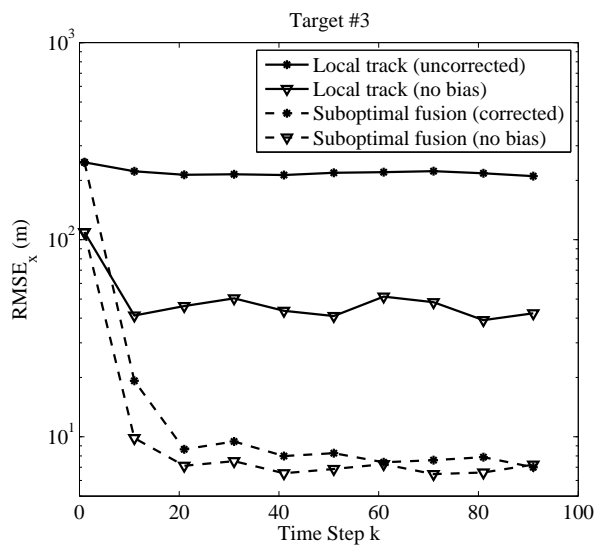}
\caption{RMSE of local track (sensor $1$) and the output of the fusion algorithm including offset biases for all sensors in logarithmic scale}
\label{fig:Fusion}
\end{figure}

To show the performance of the new algorithm in the fusion step, the results in terms of RMSE of the local track estimates for a specific sensor (sensor $1$) and the fused estimates are presented in Figure \ref{fig:Fusion} in logarithmic scale. To compare the results, the RMSE values of the local tracks and fused tracks with no biases are also included.

Figure \ref{fig:Fusion} shows that the sequential fusion step used in the proposed bias estimation algorithm (\fbe) is a viable solution to the fusion problem. Clearly, the RMSE of the fused track with corrected measurements is between those of the local track and the fused track of measurements with no biases. This observation indicates that the bias correction and fusion steps work well, which is another feature in the new algorithm, as this correction is done at the fusion center and not at the local trackers. In this case, there is no need for a feedback channel. This reduces the communication requirements.

\begin{figure}[htbp!]
\centering
\includegraphics[width=3.5in]{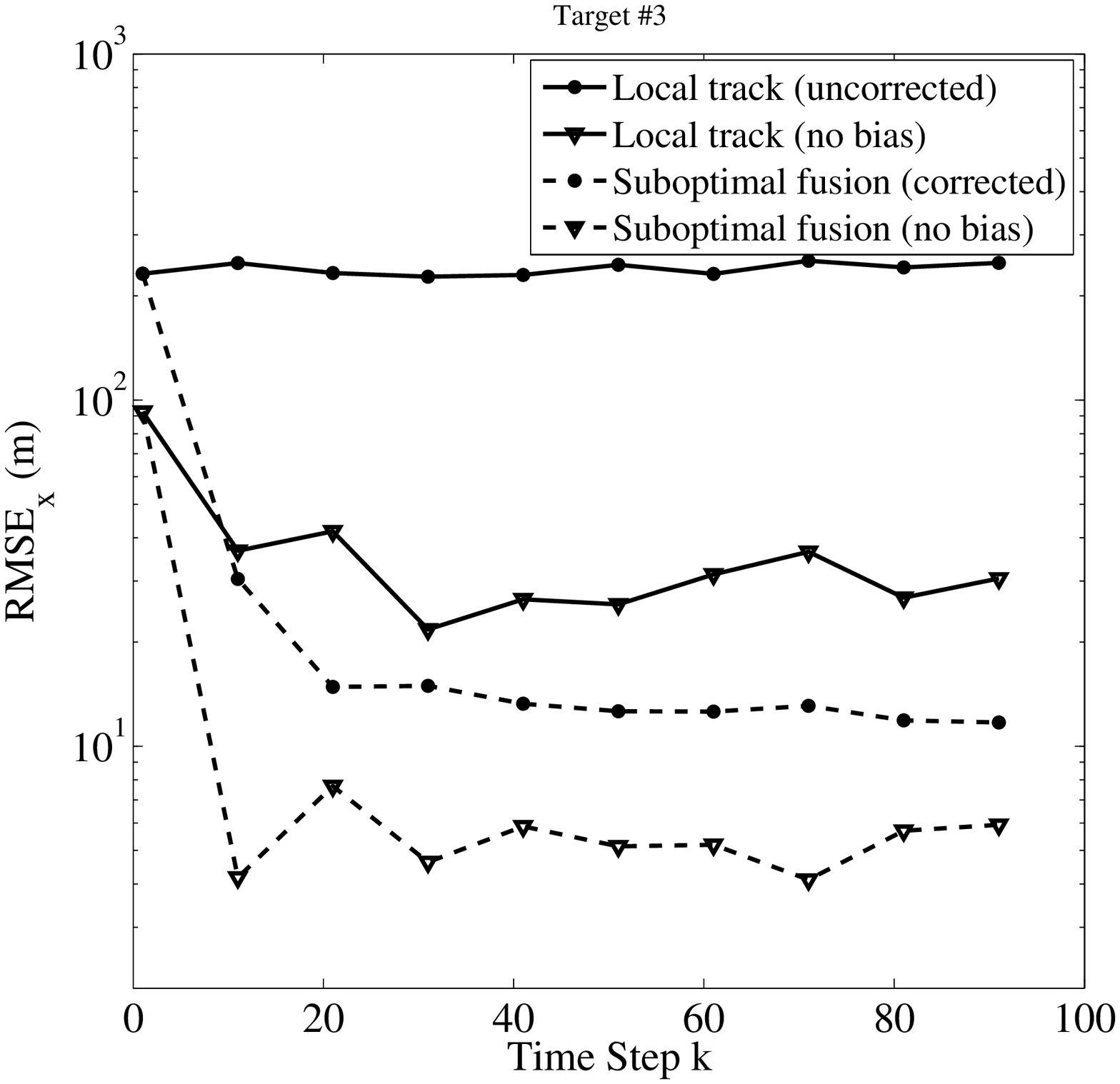}
\caption{RMSE of local track (sensor $1$) and the output of the fusion algorithm including scaling and offset biases for all sensors in logarithmic scale}
\label{fig:Fusion2}
\end{figure}

In order to further evaluate the performance of the proposed algorithm, one can assume that all sensors have scaling and offset biases. Let the value of biases be
\begin{equation}
\beta_{s}=[\begin{array}{cccc}
20\mathrm{m} & 1\mathrm{mrad} & 0.001 & 0.001\end{array}]^{\mathrm{T}}
\end{equation}
The results in terms of the RMSE of the local track estimates for a specific sensor (sensor $1$) and the fused estimates are shown in Figure \ref{fig:Fusion2} in logarithmic scale.
Figure \ref{fig:Fusion2} shows that the proposed algorithm can handle offset and scaling biases at the same time and fuse the corrected tracks in order to achieve better estimates of the targets' state.

\subsection{\label{fiveIMMCVCV} A five--sensor problem with a two--NCV IMM as local estimator}
\begin{figure}[htbp!]
\centering
\includegraphics[width=3.5in]{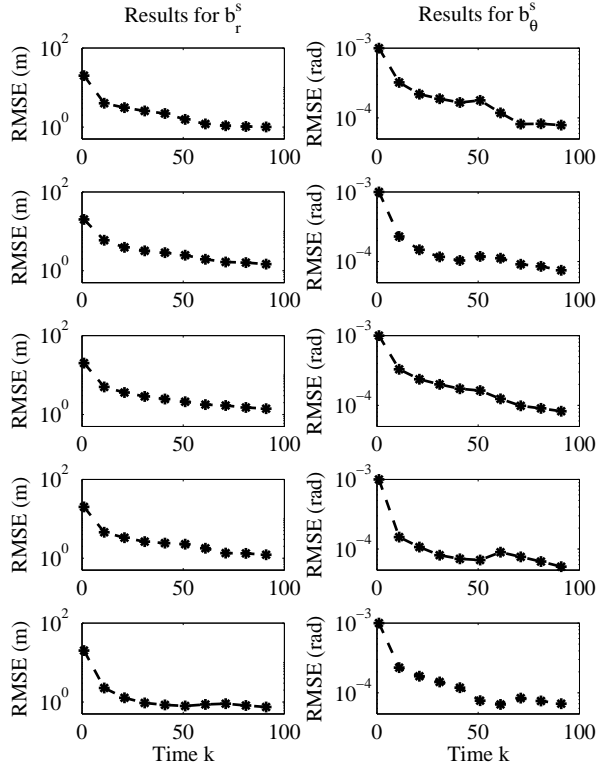}
\caption{RMSE of the bias parameter $b_{r}$ (left column) and $b_{\theta}$ (right column) for all $5$ sensors from sensor $1$ (top) to sensor $5$ (bottom) in logarithmic scale. The local trackers use IMM estimators with NCV--NCV models.}
\label{fig:5IMMCVCV}
\end{figure}

Previously, the noise levels of local tracker filters were assumed to be known. To demonstrate how well the proposed algorithm works when this information is not available, the IMM estimator is used in the next two examples as local tracker filters. To start with, an IMM estimator with two nearly constant velocity (NCV) or DCWNA Kalman filters with different noise intensities are used. The first filter uses $q_x=q_y=10\frac{\mathrm{m}^{2}}{\mathrm{s}^{3}}$ while the second one uses $q_x=q_y=2\frac{\mathrm{m}^{2}}{\mathrm{s}^{3}}$ as intensities in the east and north directions, respectively. Note that the noise intensity parameters $q_x$ and $q_y$ have the same meaning as in \cite{lin2004multisensor}, \ie, power spectral densities. To ensure accurate bias estimation, parameters of the \rlsb algorithm should be changed to handle the mismatch in the models between local trackers and fusion center which uses only an NCV model for data processing. Although there is no systematic way to select the intensities for the NCV model at the fusion center, $q$ should be the value of the higher intensity in each coordinate. Figure \ref{fig:5IMMCVCV} shows the RMSE results for the bias parameters that are estimated in the case of having NCV--NCV IMM estimators as local trackers and only one NCV model at the fusion center with inflated intensity level.

Figure \ref{fig:5IMMCVCV} shows how well the bias estimation can be handled by EXL even when it is not possible to recover the exact Kalman gains that are used the at local trackers. The difference in convergence between the previous example (Figure \ref{fig:biasRMSE}) and Figure \ref{fig:5IMMCVCV} is negligible, which shows the effectiveness of the new algorithm in different situations.

\subsection{\label{fiveIMMCACV} A five--sensor problem with nearly constant acceleration--nearly constant velocity (NCA--NCV) IMM as local estimators}
\begin{figure}[htbp!]
\centering
\includegraphics[width=3.5in]{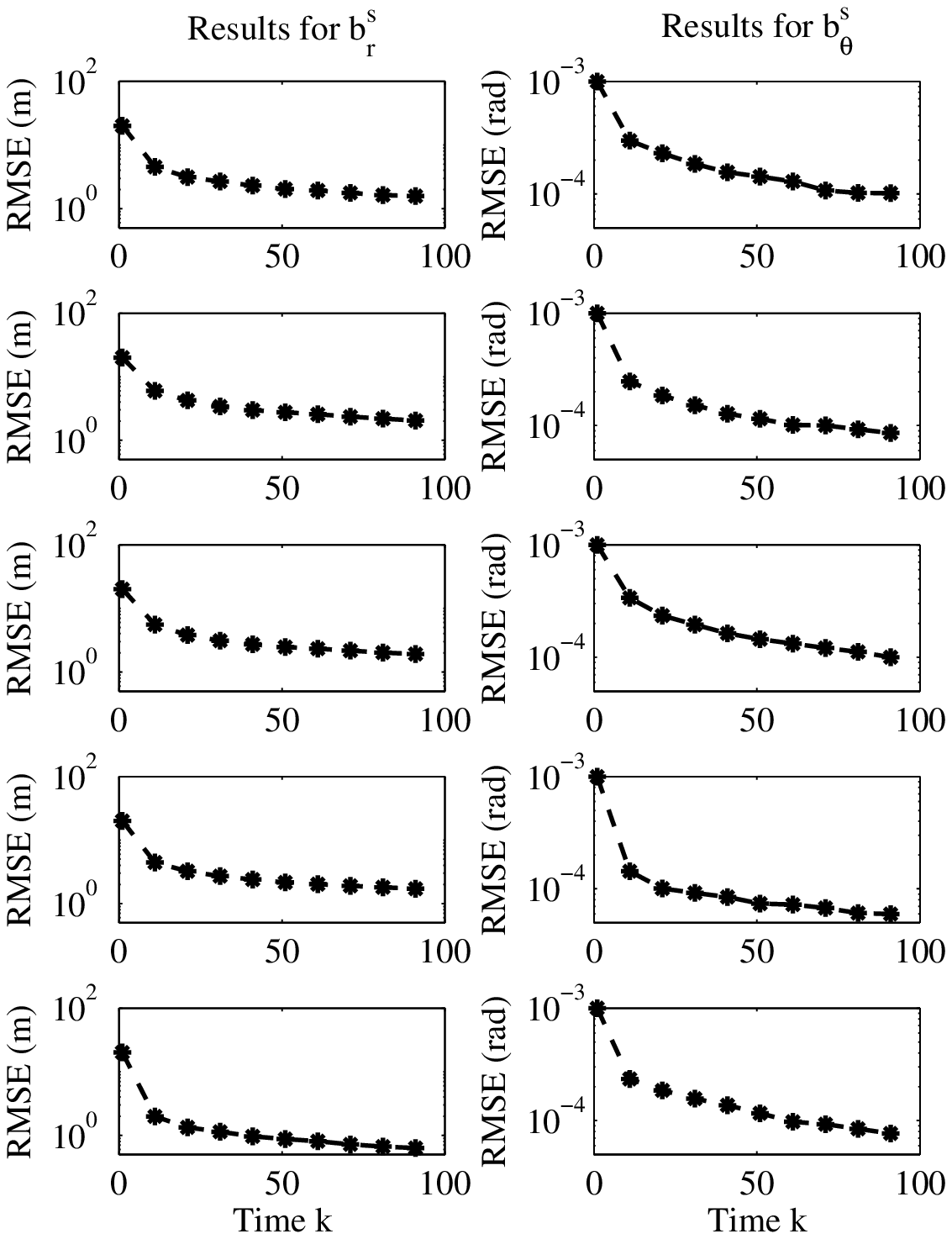}
\caption{RMSE of the bias parameter $b_{r}$ (left column) and $b_{\theta}$ (right column) for all $5$ sensors from sensor $1$ (top) to sensor $5$ (bottom) in logarithmic scale. The local trackers use IMM estimators with NCA--NCV models.}
\label{fig:5IMMAVCV}
\end{figure}

Next, we demonstrate the effectiveness of the new algorithm in recovering the Kalman gain and estimating the biases and show how well the new algorithm works in the case of mismatch in the models at the fusion center and local trackers. In this example it is assumed that local trackers are using an IMM estimator with one nearly constant acceleration (NCA) and one NCV Kalman filter with $q_x=q_y=10\frac{\mathrm{m}^{2}}{\mathrm{s}^{3}}$ for the NCA model and $q_x=q_y=2\frac{\mathrm{m}^{2}}{\mathrm{s}^{3}}$ for the NCV model. The issue in this case is that the local trackers send only the combined output from the IMM estimator without any information about the acceleration. Figure \ref{fig:5IMMAVCV} shows the results on the scenario of this subsection with $q_x=q_y=200\frac{\mathrm{m}^{2}}{\mathrm{s}^{3}}$ at the fusion center. 

Simulation results show convergence as in the previous examples (Figures \ref{fig:biasRMSE} and \ref{fig:5IMMCVCV}). This demonstrates the robustness of the new algorithm even in the presence of uncertainty about the local trackers.

\subsection{\label{pcrlb_result} Lower bound and convergence results}
The calculation of the CRLB was discussed in Section \ref{pcrlb}. Three different examples are used to demonstrate the performance of the proposed algorithm with respect to the CRLB. The first example is the scenario implemented in Subsection \ref{fiveSensorKalman}. The comparison is between the square root of the diagonal elements of the CRLB ($\sqrt{\mathrm{CRLB}\left\{ \left[\mathbf{b}\right]_{i}\right\}}$) , the square root of the diagonal elements of bias estimation covariance matrix ($\sqrt{\Sigma_{ii}}$) and the RMSE of the estimated biases. 
\begin{figure}[htbp!]
\centering
\includegraphics[width=3.5in]{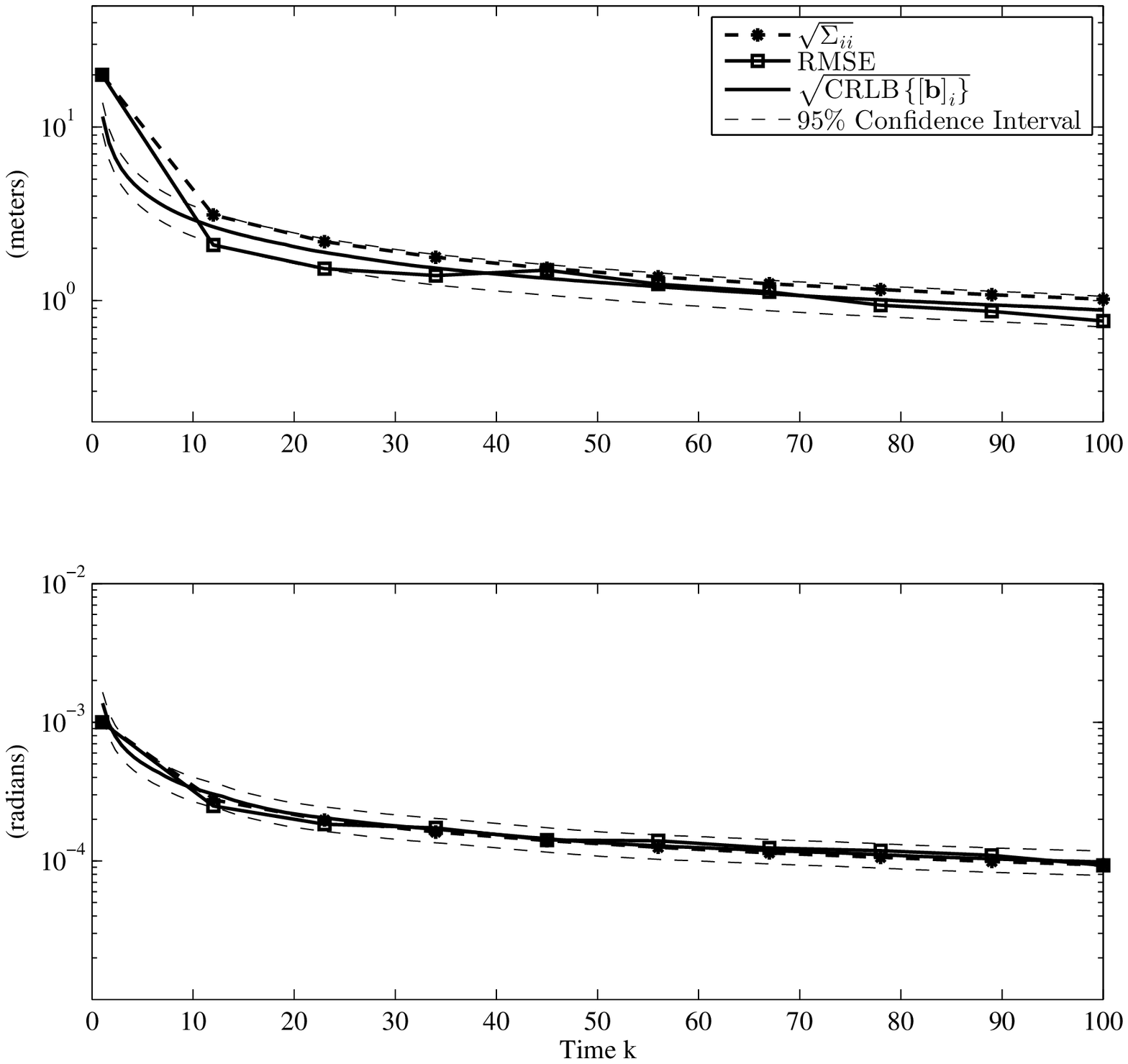}
\caption{Comparison between the square root of diagonal elements of CRLB ($\sqrt{\mathrm{CRLB}\left\{ \left[\mathbf{b}\right]_{i}\right\}}$), square root of diagonal elements of the covariance matrix of bias estimation algorithm ($\sqrt{\Sigma_{ii}}$) and RMSE of the bias estimation for the case of 5 sensors with Kalman filter as local trackers (only the results for the first sensor are shown).}
\label{fig:pcrlbKalman}
\end{figure}

\begin{figure}[ht!]
\centering
\includegraphics[width=3.5in]{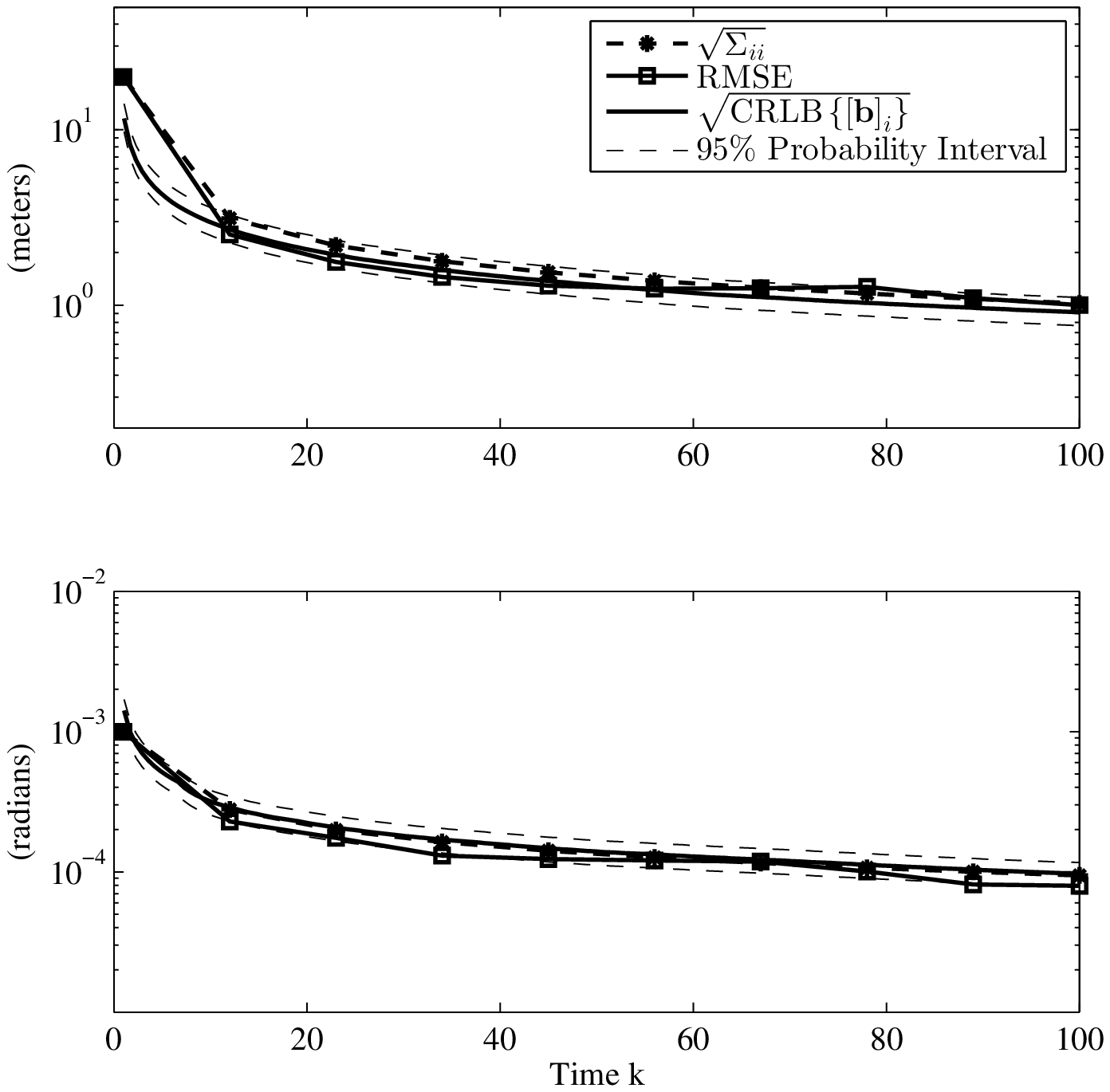}
\caption{Comparison between the square root of diagonal elements of CRLB ($\sqrt{\mathrm{CRLB}\left\{ \left[\mathbf{b}\right]_{i}\right\}}$), square root of diagonal elements of the covariance matrix of bias estimation algorithm ($\sqrt{\Sigma_{ii}}$) and RMSE of the bias estimation for the case of 5 sensors with NCV--NCV IMM estimator as local trackers (only the results for the first sensor are shown).}
\label{fig:pcrlbCVCV}
\end{figure}

\begin{figure}[ht!]
\centering
\includegraphics[width=3.5in]{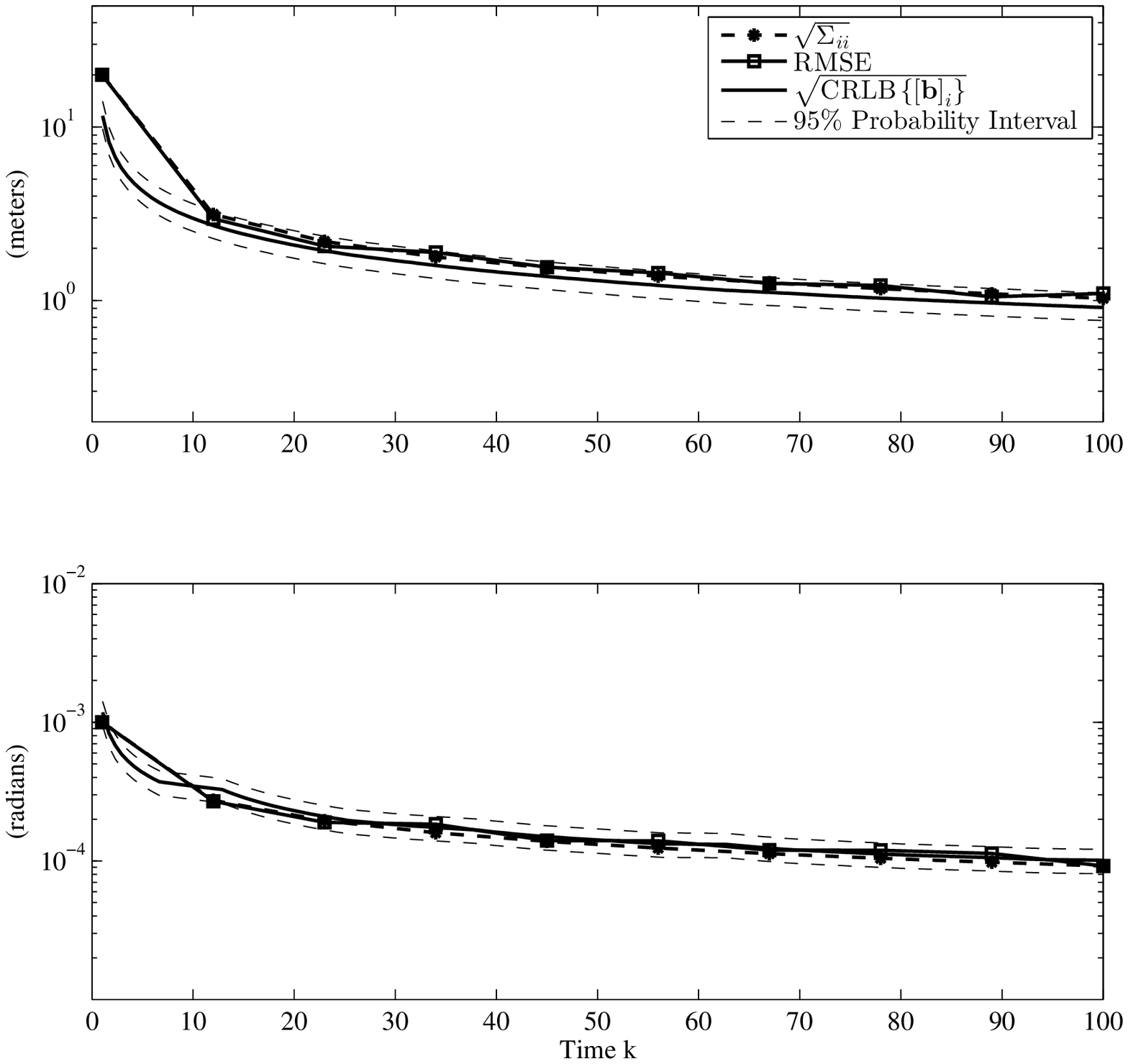}
\caption{Comparison between the square root of diagonal elements of CRLB ($\sqrt{\mathrm{CRLB}\left\{ \left[\mathbf{b}\right]_{i}\right\}}$), square root of diagonal elements of the covariance matrix of bias estimation algorithm ($\sqrt{\Sigma_{ii}}$) and RMSE of the bias estimation for the case of 5 sensors with NCA--NCV IMM estimator as local trackers (only the results for the first sensor are shown).}
\label{fig:pcrlbCACV}
\end{figure}

Figure \ref{fig:pcrlbKalman} shows that both $\sqrt{\Sigma_{ii}}$ and the RMSE follow the $\sqrt{\mathrm{CRLB}\left\{ \left[\mathbf{b}\right]_{i}\right\}}$. The results for the examples in Subsections \ref{fiveIMMCVCV} and \ref{fiveIMMCACV} are shown in Figure \ref{fig:pcrlbCVCV} and Figure \ref{fig:pcrlbCACV}, respectively. These are approximately within the $95\%$ probability region around the CRLB \cite{belfadel2013bias}. Once again, the figures show that the estimation errors follow the CRLB.

\subsection{\label{nees} Consistency of bias estimation algorithms}
\begin{figure}[ht!]
\centering
\includegraphics[width=3.5in]{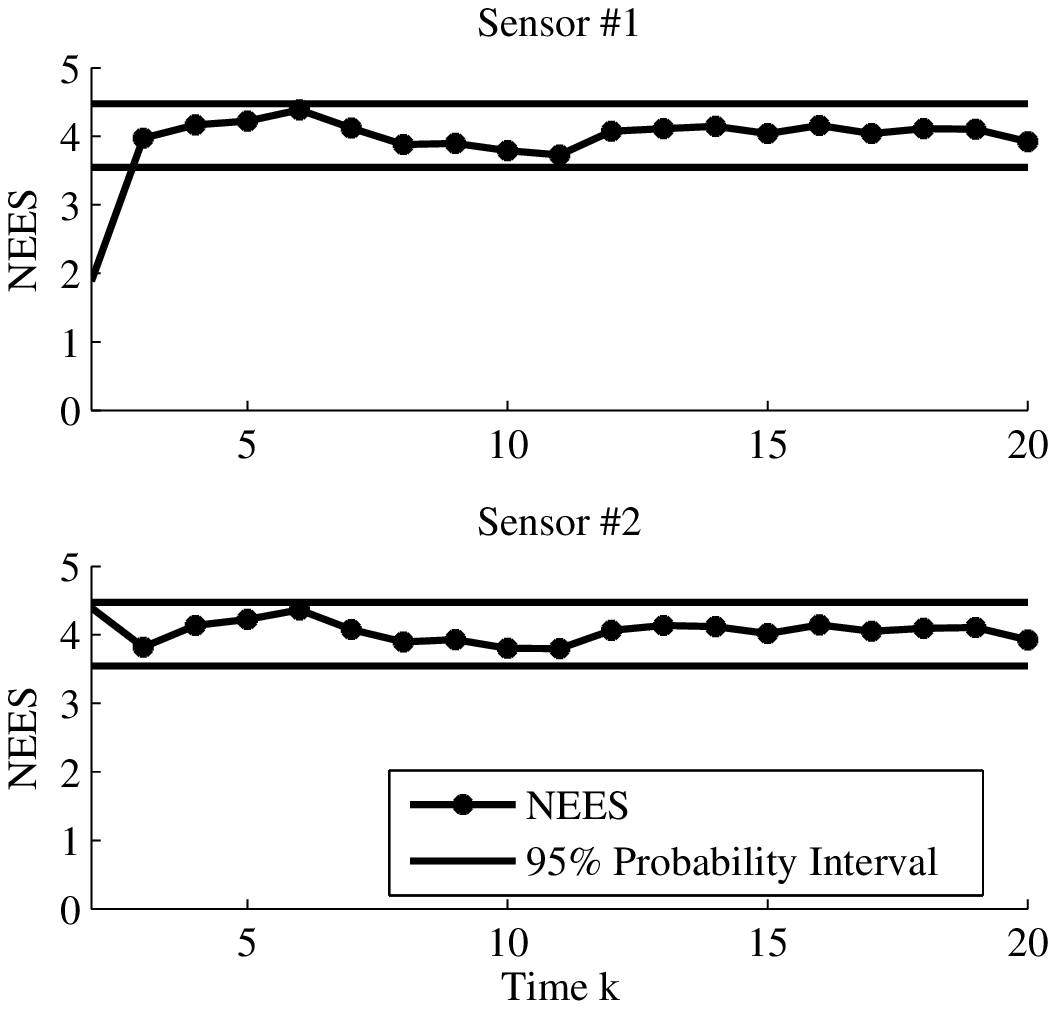}
\caption{NEES for EXL algorithm with Kalman gain recovery instead of using original Kalman gains.}
\label{fig:nees4}
\end{figure}

In this section, a brief analysis of consistency of the proposed bias estimation algorithms is given. The analysis is based on Normalized Estimation Error Squared (NEES) \cite{bar2001estimation}. First, the results for EXL algorithm are shown in Figure \ref{fig:nees4} for $100$ Monte--Carlo runs. The bounds are for the $95\%$ probability interval which shows that the EXL algorithm is consistent at each time step.

\begin{figure}[ht!]
\centering
\includegraphics[width=3.5in]{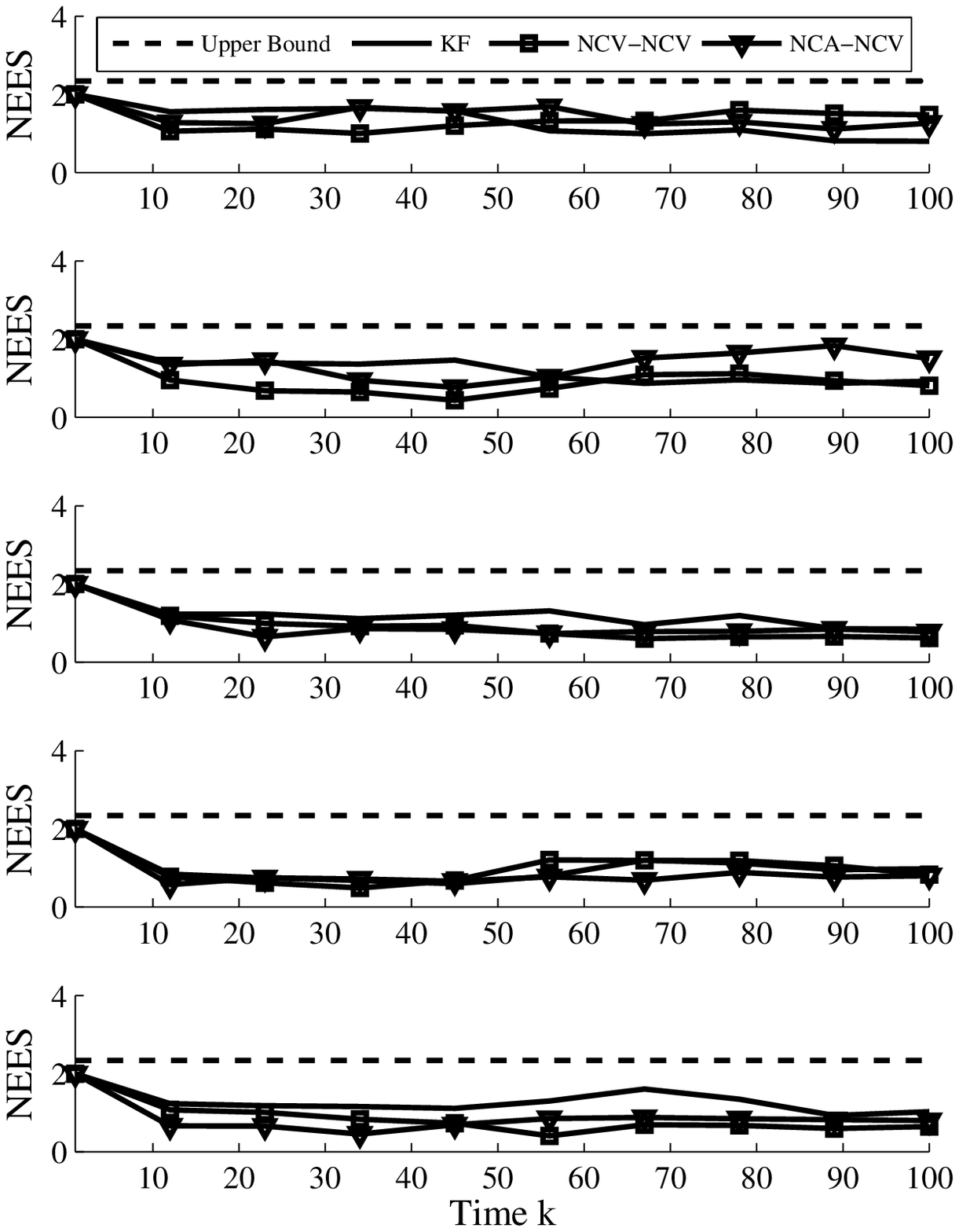}
\caption{NEES for \fbe and three different local tracker estimators (Kalman filter, NCV--NCV IMM and NCA--NCV IMM) for sensor $1$ (top) to sensor $5$ (bottom) compared to the upper--bound of $95\%$ probability interval.}
\label{fig:nees}
\end{figure}

To further examine the consistency of the proposed algorithm, the NEES for \fbe are computed and shown in Figure \ref{fig:nees} for three different types of local estimators, \ie, Kalman filter, NCV--NCV IMM and NCA--NCV IMM for $100$ Monte--Carlo runs. Here we used one--sided $95\%$ probability interval. The results show that \fbe is a pessimistic filtering approach. This is mostly due to the fact that the use of the pseudo--measurement in a Kalman filter fashion is an approximation because its error and the state prediction error at the fusion center are correlated because of the common process noise.

Finally, in Tables \ref{tab:table_r} and \ref{tab:table_theta} the RMSE, $\sqrt{\mathrm{CRLB}\left\{ \left[\mathbf{b}\right]_{i}\right\} }$
and $\sqrt{\Sigma_{ii}}$ are compared for both offset bias parameters at their last update for sensor $1$. As can be seen from Tables \ref{tab:table_r} and \ref{tab:table_theta}, both RMSE and $\sqrt{\Sigma_{ii}}$ are within the 95\% confidence region of $\sqrt{\mathrm{CRLB}\left\{ \left[\mathbf{b}\right]_{i}\right\} }$, which indicates that the parameter estimates are unbiased.
\begin{table}

\caption{\label{tab:table_r}Offset bias $\mathbf{b}_{r_{1}}$related RMSE,
$\sqrt{\mathrm{CRLB}\left\{ \left[\mathbf{b}\right]_{i}\right\} }$
and $\sqrt{\Sigma_{ii}}$ at final time-step for three different local
filters}

\centering{}%
\begin{tabular}{|c|c|c|c|}
\cline{2-4} 
\multicolumn{1}{c|}{} & Kalman filter & NCV--NCV & NCA--NCV\tabularnewline
\hline 
RMSE & $0.7621$ & $1.002$ & $1.105$\tabularnewline
\hline 
$\sqrt{\Sigma_{ii}}$  & $1.018$ & $1.025$ & $1.029$\tabularnewline
\hline 
$\sqrt{\mathrm{CRLB}\left\{ \left[\mathbf{b}\right]_{i}\right\} }$  & $0.8795$ & $0.9122$ & $0.9350$\tabularnewline
\hline 
Upper $95\%$ confidence interval & $1.055$ & $1.106$ & $1.220$\tabularnewline
\hline 
Lower $95\%$ confidence interval & $0.7036$ & $0.7666$ & $0.7481$\tabularnewline
\hline 
\end{tabular}
\end{table}
\begin{table}
\caption{\label{tab:table_theta}Offset bias $\mathbf{b}_{\theta_{1}}$related
RMSE, $\sqrt{\mathrm{CRLB}\left\{ \left[\mathbf{b}\right]_{i}\right\} }$
and $\sqrt{\Sigma_{ii}}$ at final time-step for three different local
filters}

\centering{}%
\begin{tabular}{|c|c|c|c|}
\cline{2-4} 
\multicolumn{1}{c|}{} & Kalman filter & NCV--NCV & NCA--NCV\tabularnewline
\hline 
RMSE & $9.266\times10^{-5}$ & $7.950\times10^{-5}$ & $9.197\times10^{-5}$\tabularnewline
\hline 
$\sqrt{\Sigma_{ii}}$  & $9.322\times10^{-5}$ & $9.434\times10^{-5}$ & $9.198\times10^{-5}$\tabularnewline
\hline 
$\sqrt{\mathrm{CRLB}\left\{ \left[\mathbf{b}\right]_{i}\right\} }$  & $9.826\times10^{-5}$ & $9.714\times10^{-5}$ & $10.090\times10^{-5}$\tabularnewline
\hline 
Upper $95\%$ confidence interval & $11.79\times10^{-5}$ & $11.66\times10^{-5}$ & $12.10\times10^{-5}$\tabularnewline
\hline 
Lower $95\%$ confidence interval & $7.861\times10^{-5}$ & $7.771\times10^{-5}$ & $8.069\times10^{-5}$\tabularnewline
\hline 
\end{tabular}
\end{table}

\section{\label{conclusion} Conclusions}
In this paper, a new bias estimation algorithm that works with only the state estimates and their associated covariance matrices from synchronized local trackers at varying reporting rates was presented. The algorithm does not require the stacking of the bias vectors of all the sensors together, which is a problem  for previous algorithms with a large number of sensors in the surveillance area. Also, the new algorithm  works without the local filter gains, which are not available at the fusion center in practical systems. In addition, it gives a solution to the problem of joint fusion and bias estimation. The results from simulations show that the algorithm works reliably in different scenarios with various numbers of sensors. Furthermore, this algorithm can work with low data rates between the sensors and the data fusion center. Finally, the CRLB for multisensor--multitarget scenarios with bias estimation was presented and the RMSE results matched well with the CRLB. This demonstrates the statistical efficiency and the versatility of the new algorithm.

\appendices
\section{\label{appA} Derivation of the equivalent measurement covariance \eqref{eq:TrackletP}}
Using the properties
\begin{eqnarray}
\mathbb{E}\left[\tilde{\mathbf{u}}_{s}\left(k,k'\right)\mid\mathbf{Z}^{k'}_{s}\right] & = & 0\\
\mathbb{E}\left[\tilde{\mathbf{u}}_{s}\left(k,k'\right)\mid\mathbf{Z}^{k}_{s}\right]&\neq&0
\end{eqnarray}
and
\begin{eqnarray}
\mathbb{E}\left[\tilde{\mathbf{x}}\left(k\mid k\right)\tilde{\mathbf{x}}\left(k\mid k'\right)^{\mathrm{T}}\mid\mathbf{Z}_{s}^{k'}\right] & = & \mathbb{E}\left[\mathbb{E}\left[\left[\mathbf{x}_{s}(k)-\hat{\mathbf{x}}_{s}\left(k\mid k\right)\right]\right.\right.\nonumber \\
 &  & \left.\left.\left[\mathbf{x}_{s}(k)-\hat{\mathbf{x}}_{s}\left(k\mid k\right)+\hat{\mathbf{x}}_{s}\left(k\mid k\right)-\hat{\mathbf{x}}_{s}\left(k\mid k'\right)\right]^{\mathrm{T}}\mid\mathbf{Z}_{s}^{k}\right]\mid\mathbf{Z}_{s}^{k'}\right]\\
 & = & \mathbb{E}\left[\mathbb{E}\left[\left[\mathbf{x}_{s}(k)-\hat{\mathbf{x}}_{s}\left(k\mid k\right)\right]\left[\mathbf{x}_{s}(k)-\hat{\mathbf{x}}_{s}\left(k\mid k\right)\right]^{\mathrm{T}}\mid\mathbf{Z}_{s}^{k}\right]\mid\mathbf{Z}_{s}^{k'}\right]\nonumber \\
 &  & +\mathbb{E}\left[\underset{=0}{\underbrace{\mathbb{E}\left[\left[\mathbf{x}_{s}(k)-\hat{\mathbf{x}}_{s}\left(k\mid k\right)\right]\mid\mathbf{Z}_{s}^{k}\right]}}\left[\hat{\mathbf{x}}_{s}\left(k\mid k\right)-\hat{\mathbf{x}}_{s}\left(k\mid k'\right)\right]^{\mathrm{T}}\mid\mathbf{Z}_{s}^{k'}\right]\nonumber \\
 & = & \mathbf{P}_{s}\left(k\mid k\right)
\end{eqnarray}
$\mathbf{U}_{s}\left(k,k'\right)$ can be expanded as
\begin{eqnarray}
\mathbf{U}_{s}\left(k,k'\right) & = & \mathbb{E}\left[\tilde{\mathbf{u}}_{s}\left(k,k'\right)\tilde{\mathbf{u}}_{s}\left(k,k'\right)^{\mathrm{T}}\mid\mathbf{Z}^{k'}_{s}\right]\nonumber \\
 & = & \mathbf{A}_{s}\mathbf{P}_{s}\left(k\mid k\right)\mathbf{A}_{s}^{\mathrm{T}}+\left[\mathbf{A}_{s}-I\right]\mathbf{P}_{s}\left(k\mid k'\right)\left[\mathbf{A}_{s}-I\right]^{\mathrm{T}}\nonumber \\
 &  & -\mathbf{A}_{s}\mathbf{P}_{s}\left(k\mid k\right)\left[\mathbf{A}_{s}-I\right]^{\mathrm{T}}-\left[\mathbf{A}_{s}-I\right]\mathbf{P}_{s}\left(k\mid k\right)\mathbf{A}_{s}\nonumber \\
 & = & \left[\mathbf{A}_{s}-I\right]\mathbf{P}_{s}\left(k\mid k'\right)\left[\mathbf{A}_{s}-I\right]^{\mathrm{T}}-\mathbf{A}_{s}\mathbf{P}_{s}\left(k\mid k\right)\mathbf{A}_{s}^{\mathrm{T}}+\mathbf{A}_{s}\mathbf{P}_{s}\left(k\mid k\right)+\mathbf{P}_{s}\left(k\mid k\right)\mathbf{A}_{s}^{\mathrm{T}}\nonumber \\
\end{eqnarray}
where the arguments of $\mathbf{A}_{s}(k,k')$ are dropped for clarity.
By using the property
\begin{eqnarray}
\left[\mathbf{A}_{s}-I\right]\mathbf{P}_{s}\left(k\mid k'\right) & = & \left[\mathbf{P}_{s}\left(k\mid k'\right)\left[\mathbf{P}_{s}\left(k\mid k'\right)-\mathbf{P}_{s}\left(k\mid k\right)\right]^{-1}-I\right]\mathbf{P}_{s}\left(k\mid k'\right)\nonumber \\
 & = & \mathbf{P}_{s}\left(k\mid k'\right)\left[\left[\mathbf{P}_{s}\left(k\mid k'\right)-\mathbf{P}_{s}\left(k\mid k\right)\right]^{-1}-\mathbf{P}_{s}\left(k\mid k'\right)^{-1}\right]\mathbf{P}_{s}\left(k\mid k'\right)\nonumber \\
 & = & \mathbf{P}_{s}\left(k\mid k'\right)\left[\mathbf{P}_{s}\left(k\mid k'\right)-\mathbf{P}_{s}\left(k\mid k\right)\right]^{-1}\nonumber \\
 &  & \left[I-\left[\mathbf{P}_{s}\left(k\mid k'\right)-\mathbf{P}_{s}\left(k\mid k\right)\right]\mathbf{P}_{s}\left(k\mid k'\right)^{-1}\right]\mathbf{P}_{s}\left(k\mid k'\right)\nonumber \\
 & = & \mathbf{A}_{s}\left[I-I+\mathbf{P}_{s}\left(k\mid k\right)\mathbf{P}_{s}\left(k\mid k'\right)^{-1}\right]\mathbf{P}_{s}\left(k\mid k'\right)\nonumber \\
 & = & \mathbf{A}_{s}\mathbf{P}_{s}\left(k\mid k\right)
 \label{paper_A:appendix_1}
\end{eqnarray}
$\mathbf{U}_{s}\left(k,k'\right)$ can be further simplified as
\begin{eqnarray}
\mathbf{U}_{s}\left(k,k'\right) & = & \mathbf{A}_{s}\mathbf{P}_{s}\left(k\mid k\right)\left[\mathbf{A}_{s}-I\right]^{\mathrm{T}}-\mathbf{A}_{s}\mathbf{P}_{s}\left(k\mid k\right)\mathbf{A}_{s}^{\mathrm{T}}+\mathbf{A}_{s}\mathbf{P}_{s}\left(k\mid k\right)+\mathbf{P}_{s}\left(k\mid k\right)\mathbf{A}_{s}^{\mathrm{T}}\nonumber \\
 & = & \mathbf{A}_{s}\mathbf{P}_{s}\left(k\mid k\right)\mathbf{A}_{s}^{\mathrm{T}}-\mathbf{A}_{s}\mathbf{P}_{s}\left(k\mid k\right)-\mathbf{A}_{s}\mathbf{P}_{s}\left(k\mid k\right)\mathbf{A}_{s}^{\mathrm{T}}+\mathbf{A}_{s}\mathbf{P}_{s}\left(k\mid k\right)+\mathbf{P}_{s}\left(k\mid k\right)\mathbf{A}_{s}^{\mathrm{T}}\nonumber \\
 & = & \mathbf{P}_{s}\left(k\mid k\right)\mathbf{A}_{s}^{\mathrm{T}}
 \label{paper_A:appendix_2}
\end{eqnarray}
which yields \eqref{eq:TrackletP}. Note that \eqref{paper_A:appendix_1} and \eqref{paper_A:appendix_2} are transpose of each other, but, since $\mathbf{U}_{s}\left(k,k'\right)$ is symmetric, they are equal to each other.

\ifCLASSOPTIONcaptionsoff
  \newpage
\fi



%

\bibliographystyle{unsrt}
\bibliography{references}
\begin{biography}[{\includegraphics[width=1in,height=1.25in,clip,keepaspectratio]{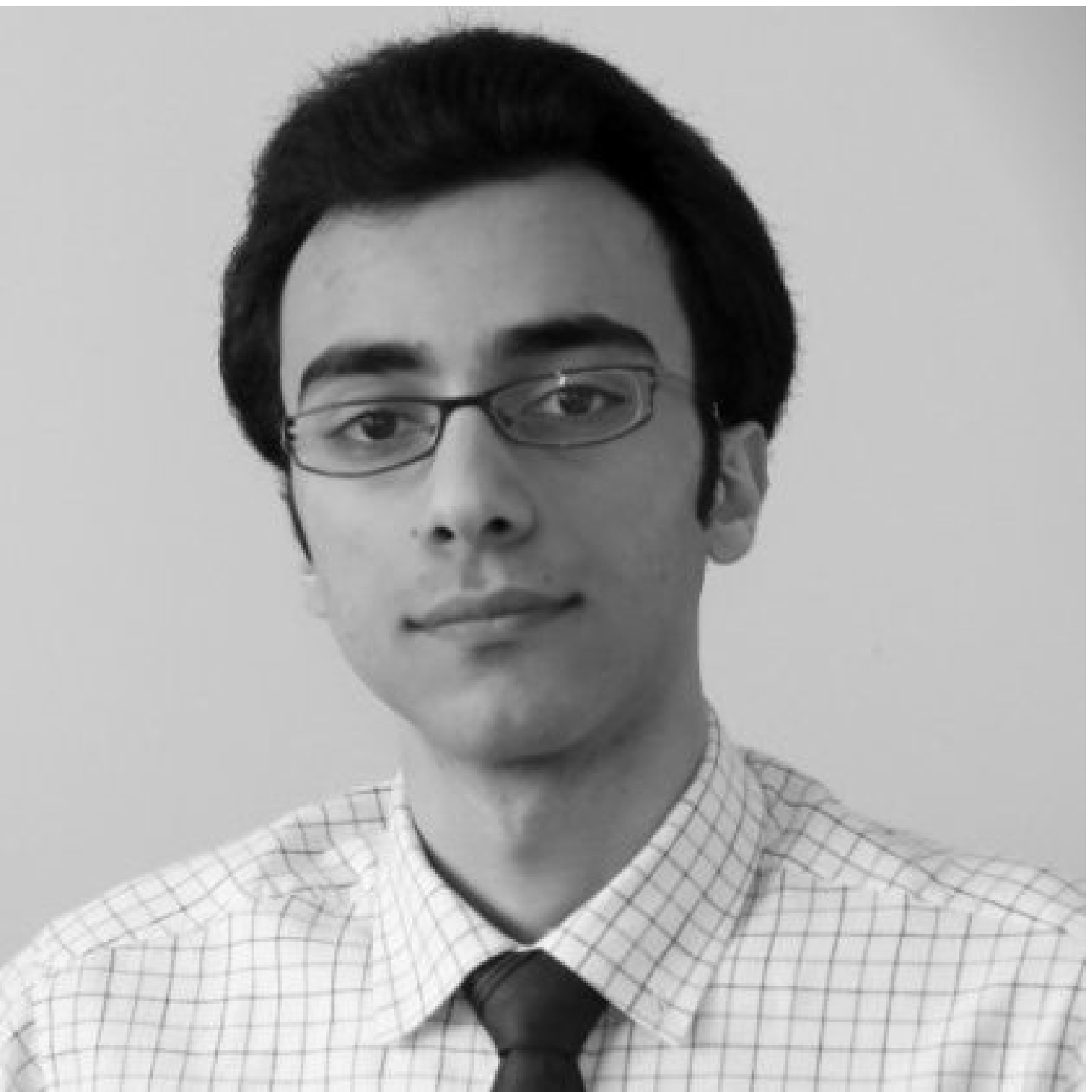}}]{Ehsan Taghavi} received the M.Sc. degree in communication engineering in 2012 from Chalmers University of Technology, Gothenburg, Sweden, where he worked on particle filter smoother. He is currently pursuing the Ph.D. degree in computational science and engineering at McMaster University, Hamilton, Canada. His research interests include estimation theory, scientific computing, signal processing, parameter estimation, mathematical modeling and algorithm design.
\end{biography}

\begin{biography}[{\includegraphics[width=1in,height=1.25in,clip,keepaspectratio]{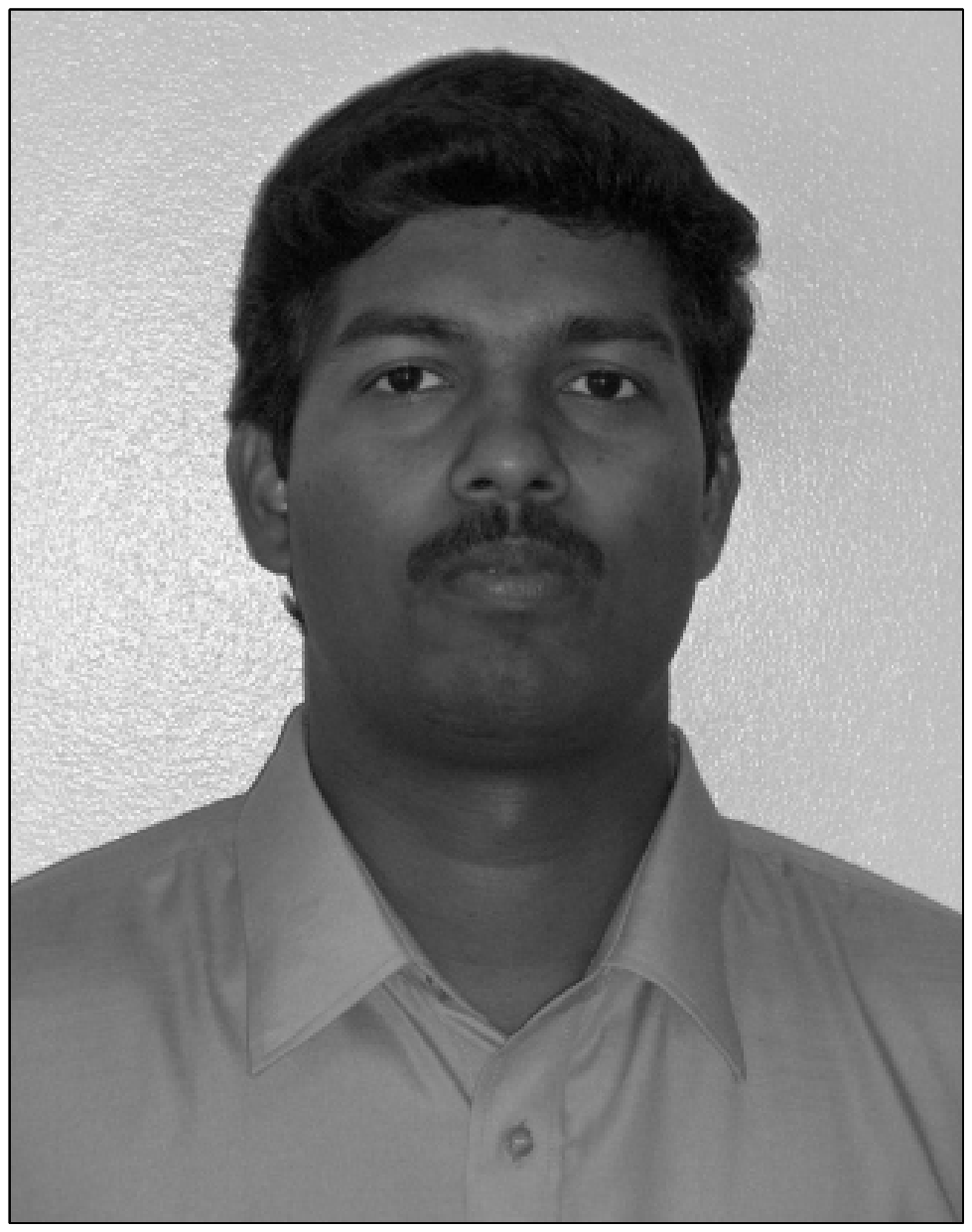}}]{Ratnasingham Tharmarasa}
was born in Sri Lanka in 1975. He received the B.Sc.Eng. degree in electronic
and telecommunication engineering from University of Moratuwa, Sri Lanka in
2001, and the M.A.Sc and Ph.D. degrees in electrical engineering from McMaster
University, Canada in 2003 and 2007, respectively.

From 2001 to 2002 he was an instructor in electronic and telecommunication
engineering at the University of Moratuwa, Sri Lanka. During 2002-2007 he was a
graduate student/research assistant in ECE department at the McMaster
University, Canada. Currently he is working as a Research Associate in the
Electrical and Computer Engineering Department at McMaster University, Canada.
His research interests include target tracking, information fusion and sensor
resource management.
\end{biography}

\begin{biography}[{\includegraphics[width=1in,height=1.25in,clip,keepaspectratio]{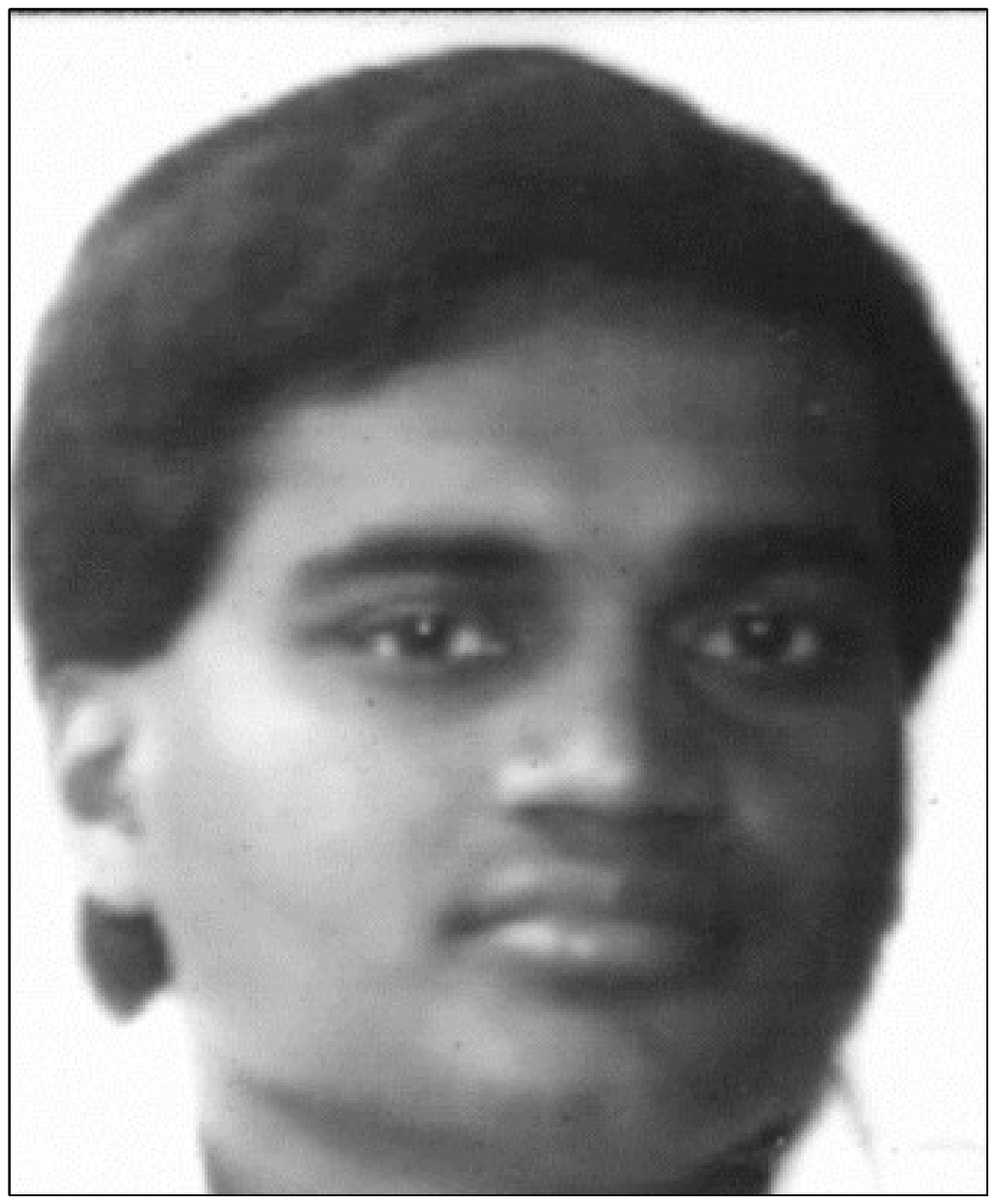}}]{Thiagalingam Kirubarajan}
(S'95–M'98–SM'03) was born in Sri Lanka in 1969. He received the B.A. and M.A.
degrees in electrical and information engineering from Cambridge University,
England, in 1991 and 1993, and the M.S. and Ph.D. degrees in electrical
engineering from the University of Connecticut, Storrs, in 1995 and 1998,
respectively.

Currently, he is a professor in the Electrical and Computer Engineering
Department at McMaster University, Hamilton, Ontario. He is also serving as an
Adjunct Assistant Professor and the Associate Director of the Estimation and
Signal Processing Research Laboratory at the University of Connecticut. His
research interests are in estimation, target tracking, multisource information
fusion, sensor resource management, signal detection and fault diagnosis. His
research activities at McMaster University and at the University of Connecticut
are supported by U.S. Missile Defense Agency, U.S. Office of Naval Research,
NASA, Qualtech Systems, Inc., Raytheon Canada Ltd. and Defense Research
Development Canada, Ottawa. In September 2001, Dr. Kirubarajan served in a
DARPA expert panel on unattended surveillance, homeland defense and
counterterrorism. He has also served as a consultant in these areas to a number
of companies, including Motorola Corporation, Northrop-Grumman Corporation,
Pacific-Sierra Research Corporation, Lockhead Martin Corporation, Qualtech
Systems, Inc., Orincon Corporation and BAE systems. He has worked on the
development of a number of engineering software programs, including BEARDAT for
target localization from bearing and frequency measurements in clutter, FUSEDAT
for fusion of multisensor data for tracking. He has also worked with Qualtech
Systems, Inc., to develop an advanced fault diagnosis engine.

Dr. Kirubarajan has published about 100 articles in areas of his research
interests, in addition to one book on estimation, tracking and navigation and
two edited volumes. He is a recipient of Ontario Premier’s Research Excellence
Award (2002).
\end{biography}

\begin{biography}[{\includegraphics[width=1in,height=1.25in,clip,keepaspectratio]{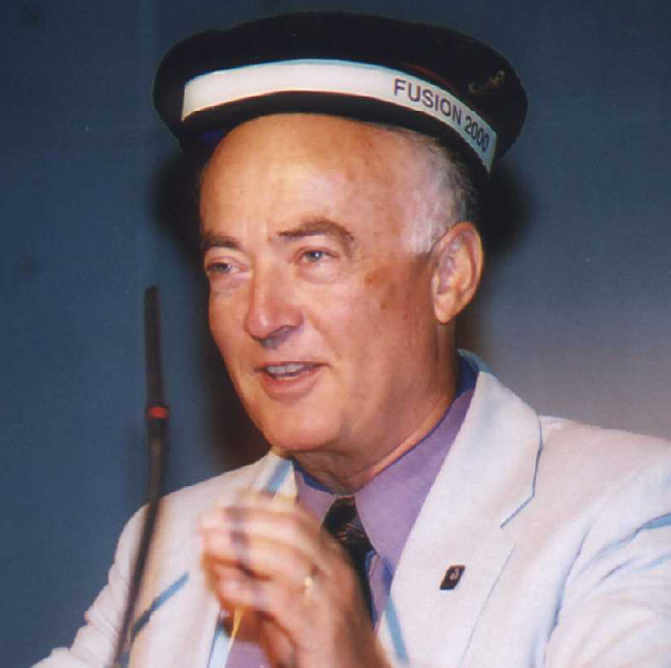}}]{Yaakov Bar--Shalom} was born on May 11, 1941. He received the B.S. and M.S. degrees from the Technion, Israel Institute of Technology, in 1963 and 1967 and the Ph.D. 
degree from Princeton University in 1970, all in electrical engineering. From 1970 to 1976 he was with Systems Control, Inc., Palo Alto, California. Currently he is Board of Trustees Distinguished Professor in the Dept. of Electrical and Computer Engineering 
and Marianne E. Klewin Professor in Engineering at the University of Connecticut. He is also Director of the ESP (Estimation and Signal Processing)  Lab. His current research interests are in estimation theory, target tracking and data fusion. He has published over 500 papers and book chapters in these areas and in stochastic adaptive control. He coauthored the monograph Tracking and Data Association (Academic Press, 1988), the graduate texts Estimation and Tracking: Principles, Techniques 
and Software (Artech House, 1993; translated into Russian, MGTU Bauman, Moscow, Russia, 2011), Estimation with Applications to Tracking and Navigation: Algorithms and Software for Information Extraction (Wiley, 2001), the advanced graduate texts Multitarget­Multisensor Tracking: Principles and Techniques (YBS Publishing, 1995), Tracking and Data Fusion (YBS Publishing, 2011),
and edited the books Multitarget­Multisensor Tracking: Applications and Advances (Artech House, Vol. I, 1990; Vol. II, 1992; Vol. III, 2000). 

He has been elected Fellow of IEEE for ``contributions to the theory of stochastic systems and of multi­ target tracking''. He has been consulting to numerous companies and government agencies, and originated the series of Multitarget­Multisensor Tracking short courses offered via UCLA Extension, at Government Laboratories, private companies and overseas. During 1976 and 1977 he served as Associate Editor of the IEEE Transactions on Automatic Control and from 1978 to 1981 as Associate Editor of Automatica. He was Program Chairman of the 1982 American Control Conference, General Chairman of the 1985 ACC, and Co­Chairman of the 1989 IEEE International Conference on Control and Applications. During 1983­87 he served as Chairman of the Conference Activities Board of the IEEE Control Systems Society and during 1987­89 was a member of the Board of Governors of the IEEE CSS. He was a member of the Board of Directors of the International Society of Information Fusion (1999--2004) and served as General Chairman of FUSION 2000, President of ISIF in 2000 and 2002 and Vice President for Publications in 2004-13. In 1987 he received the IEEE CSS Distinguished Member Award. Since 1995 he is a Distinguished Lecturer of the IEEE AESS and has given numerous keynote addresses at major national and international  conferences. He is co­recipient of the M. Barry Carlton Award for the best paper in the IEEE Transactions on Aerospace and Electronic Systems in 1995 and 2000 and recipient of the 1998 University of Connecticut AAUP Excellence Award for Research. In 2002 he received the J. Mignona Data Fusion Award from the DoD JDL Data Fusion Group. He is a member of the Connecticut Academy of Science and Engineering. In 2008 he was awarded the IEEE Dennis J. Picard Medal for Radar Technologies and Applications, and in 2012 the Connecticut Medal of Technology. He has been listed by academic.research.microsoft (top authors in engineering) as number 1 among the researchers in Aerospace Engineering based on the citations of his work.
\end{biography}

\begin{biography}[{\includegraphics[width=1in,height=1.25in,clip,keepaspectratio]{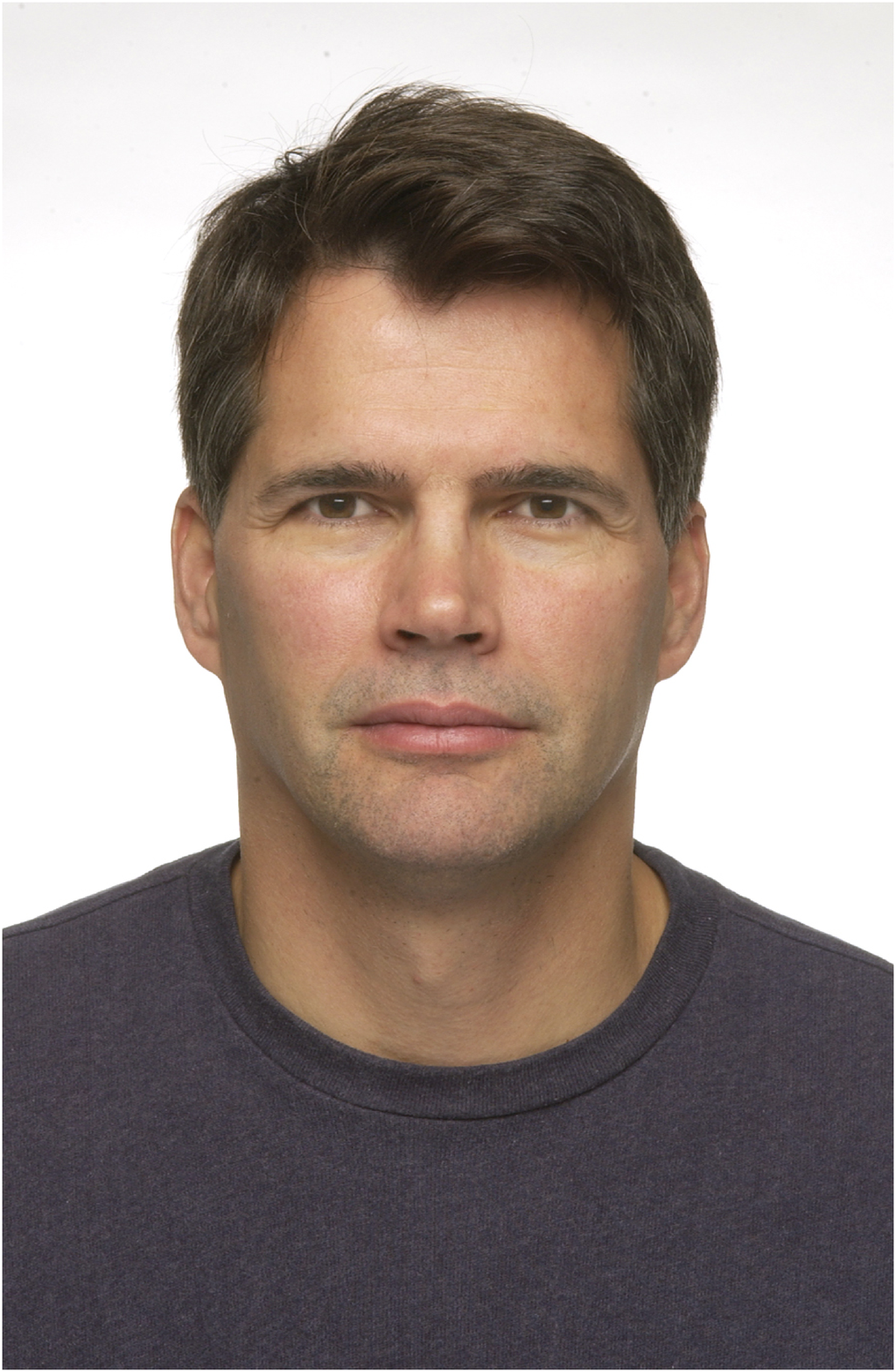}}]{Michael McDonald}
received a B.Sc (Hons) degree in Applied Geophysics from Queens University in
Kingston, Canada in 1986 and a M.Sc. degree in Electrical Engineering in 1990,
also from Queen's University.  He received a Ph.D in Physics from the
University of Western Ontario in London, Canada in 1997.  He was employed at
ComDev in Cambridge, Canada from 1989 through 1992 in their space science and
satellite communications departments and held a post-doctoral position in the
Physics department of SUNY at Stony Brooke from 1996 through 1998 before
commencing his current position as Defence Scientist in the Radar Systems
section of Defence Research and Development Canada, Ottawa, Canada. His current
research interests include the application of STAP processing and nonlinear
filtering to the detection of small maritime and land targets as well as the
development and implementation of passive radar systems.
\end{biography}
%








\end{document}